 \documentclass[twocolumn]{aastex61}

\received{February 20, 2019}
\revised{April --- , 2019}
\accepted{--- ---, 2019}

\submitjournal{ApJ}

\shorttitle{Exponential decay of Extended Emission }
\shortauthors{Kagawa et al.}

\begin{document}

\title{ Exponentially Decaying Extended Emissions Following Short
Gamma-Ray Bursts with Possible Luminosity -- E-folding Time Correlation }

\author{Yasuaki Kagawa}
\altaffiliation{JSPS Research Fellow}
\email{kagawa@astro.s.kanazawa-u.ac.jp}
\author{Daisuke Yonetoku}
\email{yonetoku@astro.s.kanazawa-u.ac.jp}
\author{Tatsuya Sawano}
\author{Makoto Arimoto}
\affil{College of Science and Engineering, School of Mathematics and
Physics, Kanazawa University, Kakuma, Kanazawa, Ishikawa 920-1192, Japan}
\author{Shota Kisaka}
\altaffiliation{JSPS Research Fellow}
\affil{Department of Physics and Mathematics, Aoyama Gakuin University,
Sagamihara. Kanagawa, 252-5258, Japan}
\affil{Frontier Research Institute for Interdisciplinary Sciences,
Tohoku University, Sendai 980-8578, Japan}
\affil{Astronomical Institute, Tohoku University, Sendai, 980-8578, Japan}
\author{Ryo Yamazaki}
\affil{Department of Physics and Mathematics, Aoyama Gakuin University,
Sagamihara. Kanagawa, 252-5258, Japan}

\begin{abstract}
 The origin of extended emissions following prompt emissions of short
 gamma-ray bursts (SGRBs) is in mystery.
 The long-term activity of the extended emission is responsible for
 promising electromagnetic counterparts to gravitational waves and, so
 that it may be a key to uncovering the progenitor of SGRBs.
 We investigate the early X-ray light curves of 26 SGRBs with known
 redshifts observed with the X-Ray Telescope aboard the {\it Neil
 Gehrels Swift Observatory} ({\it Swift}).
 We find that the exponential temporal decay model is able to describe
 the extended emissions comprehensively with a rest-frame e-folding time of
 20 -- 200 seconds.
 We also estimate the isotropic equivalent energies of the extended
 emission with the exponential decay model and of the prompt emission,
 compared with those of the prompt emission. 
 Then, it is revealed that the extended emission is 0 -- 3 orders of
 magnitude less powerful than the prompt emission.  
 We find a strong correlation between the expected maximum luminosity
 and e-folding time which can be described by a power-law with an index
 of $-3.3$ and whose chance probability of $8.2\times10^{-6}$ if there is no
 observation bias of {\it Swift}.
 The exponential temporal decay may be interpreted to come from the
 spin-down time scale of the rotation energy of a highly magnetized
 neutron star, and/or fallback accretion onto a disk surrounding a black
 hole with an exponentially decaying magnetic flux by magnetic reconnection.  
\end{abstract}


\section{Introduction} \label{sec:intro}
 
Short gamma-ray bursts (SGRBs) are a sub-class of gamma-ray bursts (GRBs)
with a duration of less than about 2 seconds.
Some SGRBs are followed by temporally extended soft X-ray emission
lasting about 100 seconds \citep{norris2006}. 
In addition, most of the extended emissions were also reported to have
comparable energy fluences with the prompt emissions \citep{perley2009,
bostanci2012}. 

SGRBs are thought to originate from a coalescence of binary compact
objects such as neutron star -- neutron star and/or black hole -- neutron
star \citep{paczynski1986,eichler1989}. In this scenario, a relativistic
jet is launched from the remnant and then powers the prompt
emission. Additionally, the merger is also expected to emit strong
gravitational waves (GWs). In fact, GW170817 observed by the LIGO-Virgo
collaboration originated from a binary neutron star merger
\citep{abbott2017}.
Therefore SGRBs and the following extended emissions are promising
electromagnetic counterparts to GW events. 
In particular, the extended emission may be more isotropic than
the beamed prompt emission because of a weak variability of the
observed light curve \cite[opening
angle of $>10^{\circ}$;][]{bucciantini2012}.
SGRB afterglows following the extended emission would be also important
for the localization of GW sources.
In general, the afterglow can be distinguished by two or more
segments such as a shallow-decay emission, so-called plateau emission
with a duration of $10^3$ -- $10^4$ seconds,
\citep{gompertz2013,gompertz2014}, and a normal-decay component,
which is thought to arise from an external shock between the
relativistic jet and the circumburst medium swept up by the jet
\citep[e.g.,][]{paczynski1993, gehrels2005}. 

The origin of the temporally extended emission is, however, still in
mystery.
This has been also debated for some cases of the merger
remnant, for example, a spin-down energy loss of a rapidly spinning
magnetar \citep{metzger2008} and/or a fallback accretion of tidally
ejected mass onto a disk surrounding a black hole \citep{barkov2011, 
kisaka2015}.
The type of the remnant depends on the equation of state of
the neutron star and the total binary mass.

In order to reveal the origin of the extended emission, some studies on
the temporal behavior were performed \cite[e.g.,][]{gompertz2014,
nathanail2015}.
From the magnetar engine model, the X-ray light curves of SGRBs can be
described by the power-law decay index of $-2$ by considering the dipole
radiation \citep[e.g.,][]{zhang2001}. The study of the light curves
following SGRBs with this decay slope have been performed
\citep[e.g.,][]{lu2015}.
\cite{kisaka2017} discussed a power-law model with a power-law index of
$-40/9$ \citep{kisaka2015} by considering the black hole engine model
with Blandford  -- Znajek jet \citep{bz1977} and ejecta fallback
\cite[e.g.,][]{rosswog2007}, and then from the observed light curve
study with this model, it was concluded that in the luminosity --
duration plane the extended emission has a different distribution from
that of the plateau emission (i.e., bimodal distribution).

On the other hand, \cite{kagawa2015} reported that the exponential
temporal decay model was able to fit some extended emissions with an
e-folding time of 50 -- 100 seconds.
Similar studies on the exponential decay has been performed for early
X-ray emissions of long GRBs (LGRBs)
\cite[e.g.,][]{willingale2007, sakamoto2007, imatani2016}.
Therefore, in this paper, we report a systematic study on
phenomenological modeling for the extended emission light curves of
SGRBs by adopting the exponential and power-law decay models,
and a comparison of both models for discussing which model is suitable
for describing the observed light curves of the extended mission.

This paper is constructed as follows.
In Section \ref{sec:reduction}, we create X-ray light curves of the
extended emission following 26 SGRBs with known redshifts.
In Section \ref{sec:lcfit},
we systematically study the temporal decay properties of the extended
emission by adopting two types models; exponential and phenomenological
power-law decay models. 
Then, in Section \ref{sec:discussion},
comparing these models, we confirm that the exponential
decay model comprehensively describes most of the extended emissions.
After that, we precisely compare the bolometric energy of the prompt
emission with that of the extended emission integrated over the
entire exponential decay model.
Then, we find a strong correlation between the expected maximum
luminosity and the e-folding time of the exponential model.
Finally, we discuss physical origins of the extended emission
represented by the exponential decay model.

\section{Spectral analysis} \label{sec:reduction}
\subsection{Event Selection} \label{subsec:selection}

First, we pick up 114 SGRBs with a duration of $T_{90}<2.0$ seconds
from the {\it Neil Gehrels Swift Observatory} ({\it Swift}) GRB
webpage\footnote{\url{https://swift.gsfc.nasa.gov/}}, detected until the
end of August 2018, where $T_{90}$ corresponds to the time interval
which contains the 90\% of the total observed photons in the 50 -- 300 keV
energy band.
We also focus on possible SGRB candidates whose $T_{90}$ is longer
than 2 seconds due to the bright extended emission episode following
the prompt initial spikes. 
From the events reported in the {\it GRB Coordinates Network} (GCN)
circulars\footnote{\url{https://gcn.gsfc.nasa.gov/}},
we include 14 possible SGRBs whose $T_{90}$ of initial spike is less than 2
seconds\footnote{GRB~050911, 061210 
080123, 080503, 090531B, 090715A, 090916, 091117, 100213A, 110402A,
130716A, 130822A, 150424A, and 160303A}.
In addition, we add 13 GRBs whose $T_{90}$ of the initial spike is
slightly larger than 2 seconds but considered to be SGRB events because
their spectral time lags are consistent with zero which is expected in
general SGRBs\footnote{GRB~051227, 060717, 100816A, 161001A,
160303A, 171103A, and 180618A} \citep[e.g.,][]{cheng1995, yi2006}, or
they have a hard spectral photon index\footnote{050724, 060614, 061006, 070714B,
090309, and 171007A}
\citep[e.g.,][]{kouveliotou1993}.
Next, we select SGRBs with known redshifts from the event list to
correct the observed flux to the isotropic luminosity in order to
investigate the intrinsic behavior.

We use the X-ray extended, plateau, and/or normal-decay emissions
data observed by the Burst Alert Telescope (BAT) and the X-Ray 
Telescope (XRT) \citep{gehrels2004} to investigate the early X-ray
properties of SGRBs.
Some extended emissions were not observed with the BAT but with only the
XRT \citep{kagawa2015, kisaka2017}. 
Thus, in order to discuss the extended emissions comprehensively, it is
necessary to analyze the X-ray data of SGRBs.
Since extended emissions are thought to last $\sim 100$ seconds,
we exclude 4 events observed by the XRT over 300~seconds since the
triggers\footnote{GRB~061210, 071010B, 150101B, and 170428A}.
Furthermore, we refer UK {\it Swift} Science Data
Center\footnote{\url{http://www.swift.ac.uk/index.php}}$^,$
\footnote{\url{http://www.swift.ac.uk/xrt_curves}} which shows the quick
look data of observed SGRBs obtained by the automatic analysis
\citep{evans2007, evans2009}. 
We also reject 7 dim bursts\footnote{GRB~060502B, 061217, 070429B,
071112, 100206A, 140622A, and 141212A} whose X-ray light curves of the
XRT data consist of 4 or less data points because a light curve fitting
model described in Section \ref{sec:lcfit} has 5 free parameters.
After these event selections, we obtain 26 SGRBs with known redshifts as
listed in Table \ref{table:grblist} from the 141 {\it Swift} SGRBs
(including SGRB candidates). 

\subsection{Light Curve Creation based on Spectral Analysis for the
  XRT Data} \label{subsec:xrtsp} 
First, we extract X-ray signals within an image region of 
$40 \times 30$ rectangular pixels with a rotation angle along the
spacecraft attitude for windowed timing (WT) mode data, and 20 pixels in radius
(corresponding to $\sim$ 47 arcsec) for photon counting (PC) mode data. 
These are recommended region sizes described in the {\it Swift}/XRT
software guide version 1.2\footnote{\url{https://swift.gsfc.nasa.gov/analysis/}}.
We also extract a background signal from the image region without any
X-ray sources (under the sensitivity of the XRT). 
The region size is a rectangular with $30 \times 30$ pixels for WT mode,
and a circle as large as possible (at least 20 pixels) for PC mode
data. The source and background regions do not overlap with each other.

After that, we perform time-resolved spectral analysis and estimate
precise energy flux of the extended emission.
This is because the extended emission generally shows rapid spectral
softening \citep{kagawa2015}, and a common method of conversion from
photon flux to energy flux with average spectral parameters can not be
adopted for the extended emission.
Then we extract time-resolved spectra from WT and PC mode data of the
selected SGRBs. 
In order to conserve an uniform statistical uncertainty for each
spectrum, we divide the entire data into several time bins to keep the
same number of photons (about 256 photons for WT mode, and 128 photons
for PC mode, respectively) inside each time bin.

We consider a single power-law spectral model considering Galactic and
extra-galactic photo-electric absorptions (``phabs'' and ``zphabs''
model, respectively). 
The exact formula of the model is
\begin{eqnarray}
\label{eq:power-law}
N(E) =  
 e^{-N_{{\rm H}}^{{\rm Gal}}\sigma(E)} \times e^{-N_{{\rm H}}^{{\rm
 ext}}\sigma((1+z)E)} \times K \Bigr( \frac{E}{1~{\rm keV}} \Bigl)^{-
 \Gamma}.
\end{eqnarray}
Here, $N(E)$ is in units of $\rm{photons~cm^{-2}~s^{-1}~keV^{-1}}$.
$N_{{\rm H}}^{{\rm Gal}}$ and $N_{{\rm H}}^{{\rm ext}}$ are the Galactic
and extra-galactic hydrogen column density, in units of $10^{22}~{\rm
atoms~cm^{-2}}$, respectively. $\sigma(E)$ is a photo-electric
cross-section (not including Thomson scattering) and $z$ is a redshift.
A quantity $K$ is a normalization of power-law model of 1 keV
and $\Gamma$ is a photon index.
We perform spectral fitting with this model for the time-resolved
spectra using XSPEC version 12.10.0 \citep{arnaud1996}, and obtain the best-fit parameters. the results are shown
in Figure~\ref{fig:spectra}. 
The photon absorption below $\sim 1$ keV may affect the flux estimate.
Thus, to suppress the uncertainty of the absorption we adopt an energy
band of 2/$(1+z)$ -- 10/$(1+z)$ keV corresponding to the 2 -- 10 keV
band in the rest frame.

Additionally, we perform a time-averaged spectral analysis for the PC
mode data to investigate the light curve of dim events in the selected
SGRBs.
First, we make a light curve of photon count rate in which each time bin
contains 25 photons at least to keep the statistical uncertainty.
Then, we make an integrated spectrum during the same epoch of the
created light curve.
We can obtain the time-averaged energy flux and spectral parameters,
and also estimate a conversion factor from the average photon flux to
the averaged energy flux.
After that, we create the light curve in energy flux with the conversion
factor on the assumption that the spectral parameters are stable during
the focusing epoch. 
The obtained the light curve of the energy flux with time-averaged
spectral analysis are also shown in Figure~\ref{fig:spectra}.
In this work, for the PC mode data, we adopt the time-averaged spectral
analysis.

%
%

\subsection{{\it Swift}/BAT Detection} \label{subsec:batsp}
Since the extended emission in GRB~050724, 060614, and 070714B 
are bright, and detected with the BAT, we are able to perform a
single power-law fit for the energy spectra of the {\it Swift}/BAT data.
We show the fitting results of the energy flux and photon index in 
Figure \ref{fig:spectra}.
Here, for the BAT data, the energy flux is extrapolated to
2/$(1+z)$ -- 10/$(1+z)$ keV band from the data of 15 -- 150 keV in the observer frame.

For the other bursts without significant detection by the BAT, we
give a detection limit of the BAT.
Extrapolating the 5$\sigma$ sensitivity curve of $3.0 \times 10^{-8}
\times \sqrt{T/(1\ {\rm s})}$~erg~cm$^{-2}$~s$^{-1}$ (in 15 -- 150 keV)
\citep{lien2016}, 
we provide the limit in the energy band of 2/$(1+z)$ -- 10/$(1+z)$ keV.
Here, $T$ is a integration time in the observer frame since the burst
trigger in units of second, and we assume that the photon index of the
energy spectrum is 2, which is an averaged value of the three events
detected with the BAT as shown in Figure \ref{fig:spectra} and
consistent with the results of previous works \citep[e.g.,][]{kaneko2015,lien2016}.
This sensitivity curve is adopted in the fitting analysis of the light
curves (Section \ref{subsec:fitresult}) and the discussion about the
suitable model for the extended emission light curve (Section \ref{subsec:disexp}).

\section{Light Curve Modeling} \label{sec:lcfit}

\subsection{Exponential and Power-law Decay Models} \label{subsec:model}
In \cite{kisaka2017}, the extended emissions were systematically
investigated with a phenomenological power-law decay model. 
An exponential decay model was reported to be also 
acceptable to describe some extended emissions in \cite{kagawa2015}.
In this paper, we study the X-ray light curves of the selected SGRBs for
the extended emission components with two models: the exponential (EXP)
model and phenomenological power-law (PL) model.
Both models contain the phenomenological power-law decay component to 
describe the following plateau and/or normal-decay emission episodes
\citep{kisaka2015}. The exact light curve models are as follows; 
\begin{eqnarray}
\label{eq:exp-lc}
L(t) =  
L_{{\rm EE}}~{\rm exp}(- t/ \tau_{{\rm EE}}) +
L_{{\rm pla}}\Bigr(1 + \frac{t}{T_{{\rm pla}}} \Bigl)^{-\alpha_{{\rm pla}}},
\end{eqnarray}
for the EXP model, and 
\begin{eqnarray}
\label{eq:exp-lc}
L(t) =  
L_{{\rm EE}}\Bigr(1 + \frac{t}{T_{{\rm EE}}} \Bigl)^{-\alpha_{{\rm EE}}} + 
L_{{\rm pla}}\Bigr(1 + \frac{t}{T_{{\rm pla}}} \Bigl)^{-\alpha_{{\rm pla}}},
\end{eqnarray}
for the PL model, respectively.
Here, $t$ is a rest-frame time since the burst trigger, and an isotropic
luminosity $L(t)$ is in units of erg~s$^{-1}$.
Parameters $L_{{\rm EE}}$,  $T_{{\rm EE}}$, $L_{{\rm pla}}$ and
$T_{{\rm pla}}$ are the normalization of isotropic luminosity and
the rest-frame durations of the extended and plateau emissions since the burst
trigger, respectively. 
Parameters $\alpha_{{\rm EE}}$ and $\alpha_{{\rm pla}}$ are temporal
indices of the extended and plateau emissions, respectively.
In EXP model, $\tau_{{\rm EE}}$ is an e-folding time in the rest frame of SGRBs. 
These two functions are referred from \cite{yamazaki2009, kagawa2015,
kisaka2015, kisaka2017}.

In this work, we systematically perform temporal fitting with both EXP
and PL models, and compare the fitting results.
Here, we consider three cases,  
(I) EXP model with the free parameter $\alpha_{{\rm pla}}$, 
(II) PL(BH) model with the fixed parameters $\alpha_{{\rm EE}}$, and
$\alpha_{{\rm pla}}$ as 40/9 $\sim$ 4.44,
and an additional case of (III) PL(MG) model with the fixed parameters 
$\alpha_{{\rm EE}}$ and $\alpha_{{\rm pla}}$ as 2.
The value of 40/9 in case (II) is derived in \cite{kisaka2015}
by considering the black hole engine model with Blandford - Znajek jet
\citep{bz1977} and ejecta fallback \cite[e.g.,][]{rosswog2007}.
Case (III) corresponds to the dipole spin-down formula usually
considered in magnetar model \citep[e.g.,][]{zhang2001}.
Note that we discuss only the extended emission component in this paper.

\subsection{Fitting Results} \label{subsec:fitresult}

In Figure \ref{fig:lcfit}, we show X-ray light curves in terms of
isotropic luminosity of the selected SGRBs estimated in Section
\ref{subsec:xrtsp} and \ref{subsec:batsp} and also the best-fitted model
functions. Here, to convert energy flux to isotropic luminosity, we use
cosmological parameters of Hubble constant $H_0 = 67.4$
km~s$^{-1}$~Mpc$^{-1}$, matter density $\Omega_{m} = 0.315$, and dark
energy density $\Omega_{\Lambda} = 0.685$ \citep{aghanim2018}.

First, we evaluate whether the flux expected from the EXP model is
consistent with the BAT detection limit described in Section \ref{subsec:batsp}.
For GRB~090510 and 100816A, due to poor statistics at the early
observation phase,
we could not constrain the fitting parameters such as $L_{{\rm EE}}$ and
$\tau_{{\rm EE}}$. 
Therefore, for these events, we set the maximum value of $L_{{\rm EE}}$
as an upper limit when we assume the EXP model does not exceed
the BAT detection limit curve.
For GRB 080123 and 150424A, there were bright X-ray sources in the field
of view of the BAT and/or unexpected background fluctuations were
observed as reported in \cite{lien2016}. 
Thus, we allow these GRBs to be included in this analysis as an
exceptional case that the best-fit EXP curve of these events exceed the
BAT detection limit. This is further described in Section \ref{subsec:batsp}.

We summarize the best-fit parameters of light curve fitting in Table
\ref{table:results}. In this table, the results of plateau emission
components are not given because the parameters for most SGRBs are not
precisely determined. 
In Section \ref{subsec:disexp}, we discuss which model is suitable for
explaining the extended emission.

\subsection{Estimate of Isotropic Equivalent Energy} \label{subsec:energy}
\cite{kisaka2017} showed a comparison of the isotropic energies of the
prompt and the temporally extended emissions. In some works, the
fluences of these emission were also compared, where both emissions data
are observed with only the BAT \citep{perley2009, bostanci2012}.
On the other hand, the energy spectra of GRBs are generally well
described by a power-law function with exponential bending, so-called
the Band function \citep{band1993}, or a power-law function with
exponential cutoff \cite[e.g.,][]{sakamoto2005}. 
In the $\nu F_{\nu}$ spectrum, the bending energy, called peak energy
$E_{{\rm peak}}$, is the most intense and typically in 200 -- 300 keV
\citep{kaneko2006}.
Therefore, for the energy band of the {\it Swift}/BAT (15 -- 150 keV), it
is too narrow to measure the bolometric energy of the prompt emission,
and the previous works may underestimate the isotropic energy of the
prompt emissions as described in \cite{kisaka2017}.

In order to estimate the precise bolometric energy of the prompt
emission, we use the data of 15 events coincidently detected with the
{\it Swift}/BAT and other detectors with wide energy range,
WIND/Konus \citep{aptekar1995},
{\it Suzaku}/Wide-band All-sky Monitor (WAM) \citep{yamaoka2005},
and/or {\it Fermi}/Gamma-ray Burst Monitor (GBM) \citep{meegan2009}.
The isotropic equivalent energy of the prompt emission, $E_{{\rm iso,
pro}}$, is calculated from the fluence of SGRBs which consists of only
the initial spike component as reported in the GCN circular.
We show the events and their prompt energies in Table~\ref{table:prompt}.

In the case of the extended emission components, the estimate of the
isotropic energy depends on the light curve model.
The most extended emissions are not observed in the BAT energy range,
and their energy spectra have not been measured.
Thus, we compare the bolometric prompt emission energy with the one of
the extended emission in 2 -- 10 keV.
For the EXP and PL models, the isotropic equivalent energy $E_{{\rm iso,
EE}}$ in the rest frame energy band of 2 -- 10 keV is provided with 
\begin{eqnarray}
E_{{\rm iso, EE}} = \int^{\infty}_{0} L_{{\rm EE}}{\rm
 exp}(-t/\tau_{{\rm EE}})dt = L_{{\rm EE}}\times\tau_{{\rm EE}},
\label{eq:exp}
\end{eqnarray}
and $\frac{L_{\rm EE}T_{EE}}{\alpha_{{\rm EE}}-1}$ \citep{kisaka2017},
respectively.
In Section \ref{subsec:prompt}, we show the comparison of the isotropic
energies of the prompt and extended emissions.

\section{Discussion} \label{sec:discussion}
We systematically analyzed the early X-ray decay properties of the
selected SGRBs with known redshifts observed by the {\it Swift} satellite.
In this section, we discuss the suitable model for the temporally extended
emission and its physical origin.
\subsection{Comparison of the Temporal Decay Models}
\label{subsec:disexp}
As shown in Figure \ref{fig:lcfit},
the EXP model curve looks consistent with the observed light curve of
all the selected SGRBs.
In particular, for the three brightest events in the XRT energy band
(GRB~050724, 060614, and 160821B), the EXP model is better than the
PL(BH) model whose decay slope of $\alpha_{{\rm EE}}(=40/9\sim 4.44)$ is
not steep enough to follow the rapid decay of the extended emissions.
We note that for GRB~050724, the PL(BH) model is also acceptable because
the data points near 1000 s is well followed. However, as shown in
Figure 1 in \cite{campana2006}, a flaring activity was clearly detected
at $\sim$ 1000 s, and the data hump is not the extended emission
component. Therefore we conclud the EXP model which clearly follows
the rapid decay light curve and is better than the PL(BH) model.

Then, to systematically compare the EXP and PL(BH) models, we show a
scatter plot on the reduced $\chi^2$ of these models ($\chi^2_{\nu, {\rm
EXP}} - \chi^2_{\nu, {\rm PL(BH)}}$ plane) in Figure \ref{fig:chi} (A).
For events with $\chi^2_{\nu}<2$, both the models can be almost equally
accepted because of $\chi^2_{\nu, {\rm EXP}} \sim \chi^2_{\nu, {\rm
PL(BH)}}$.
However, for events with $\chi^2_{\nu}>2$, the EXP model is favored
because of $\chi^2_{\nu, {\rm EXP}} < \chi^2_{\nu, {\rm PL(BH)}}$ for most
events.
For the four events with $\chi^2_{\nu, {\rm EXP}} > \chi^2_{\nu, {\rm
PL(BH)}}$, although at first glance these events favor the PL(BH) model,
two of the four events (GRB~070724 and 100117A) apparently reject the
PL(BH) model
because the light curves extrapolated earlier violate the the BAT
detection limit (see Figure \ref{fig:lcfit}).
In addition, for one of the four events, GRB~160624A, the obtained
best-fitted curve of both the models is not statistically significant
because of less number of the flux points (e.g., the degrees of
freedom of the fitting results for the EXP and PL(BH) models are only 1
and 2, respectively, as listed in Table \ref{table:results}). 
Eventually, we find that only a event (GRB~051221A) significantly favors
the PL(BH) model. Thus, we conclude that in order to explain the
extended emission light curve comprehensively, the EXP model is favored. 

We also show the results of PL(MG) model fitting in Figure
\ref{fig:lcfit} and the best fit parameters are listed in Table
\ref{table:results}, where we use events whose reduced $\chi^2$ of the
PL(MG) model ($\chi^2_{\nu, {\rm PL(MG)}}$) is smaller than 7. 
Here, we do not discuss the PL(MG) model for GRB~061201 and 130603B
because $L_{{\rm EE}}$ in PL(MG) model are not determined well, although
the reduced $\chi^2$ shows that this model is acceptable.
Then, we show the scatter plot on $\chi^2_{\nu, {\rm EXP}} -
\chi^2_{\nu, {\rm PL(MG)}}$ plane in Figure \ref{fig:chi} (B).
The observed extended emission systematically prefers the EXP model over
the PL(MG) model, although only one event, GRB~051221A, clearly favors
the PL(MG) model.
Note that as is the case with the PL(BH) model, for five events such as
GRB~090510, 100816A, 160624A, the best-fit model curves exceed the BAT
detection limit and the PL(MG) model is fully rejected for the five
events.

After all, we argue that for the EXP model, it is reasonable to describe
the early X-ray light curve of 23 of the 24 selected SGRBs, where
GRB~090510 and 100816A are not included. 
As shown in Table \ref{table:results}, the e-folding times of the
temporal decay are 20 -- 200 seconds and the $L_{\rm EE}$ in the
rest-frame energy band of 2 -- 10 keV is less than $\sim 10^{50}$ 
erg~s$^{-1}$.
For both the PL(BH) and PL(MG) models, it is hard to explain the
observed light curve of the extended emission comprehensively.

\subsection{Comparison with Prompt Emissions} \label{subsec:prompt}
We show the equivalent isotropic energies of the prompt and the extended
emissions in Figure \ref{fig:eiso} and Table \ref{table:prompt},
considering that the extended emissions are described by the EXP
model. Both isotropic energies are estimated in Section 
\ref{subsec:energy}. 
As shown in Figure \ref{fig:eiso}, the extended emission has energies
smaller by a factor of 0.001 -- 1 than those of the prompt emission.
Previous works showed that the time-averaged flux of the extended emissions
is brighter than that of the prompt emission
\citep{perley2009,bostanci2012}, and the isotropic energies of these
emissions are roughly comparable \citep{kisaka2017}.
Thus the obtained result in this paper is different from that of the
previous studies. This is because we include dim GRB events detected
only with the XRT that were not included for the previous works and use
more precise values of the $E_{{\rm iso,pro}}$ than those in previous
works.

In this paper, we use the extended emission data in the 2 -- 10 keV
energy band in the rest frame as described in Section
\ref{subsec:xrtsp}, which is different from that of the previous work 
\citep[0.3 -- 10 keV;][]{kisaka2017}.
Note that for events with $z>1$ (GRB~111117A and 160410A), the energy
flux and also isotropic energy may have an uncertainty due to the
unabsorbed power-law spectrum analysis avoiding the photon absorption
below 1 keV in the observer frame.
Then, we consider an under estimate of the isotropic energy of
the extended emission caused by difference between 2 -- 10 keV and
0.3 -- 10 keV.
In Figure \ref{fig:gamma}, we show a histogram of the photon index
$\Gamma$ obtained by performing the time-resolved analysis for the WT
mode data, where we assume that the WT mode observed only the 
extended emission component.
The photon indices of the extended emission are typically $1-2$, and
Figure \ref{fig:spectra} (e.g., GRB~050724, 060614, 150424A) shows the
decay phase with $\Gamma>2$ are almost in dimmer phase of the decaying
extended emission which hardly contributes to the energy estimate.
The ratio of these fluxes in 2 -- 10 keV to 0.3 -- 10 keV with the
photon index of 1 -- 2 is at least $\sim 0.5$, and so that the
underestimate hardly affects our conclusion.
Note that the XRT can hardly observe the early flat phase of the extended
emission and the extended emission has spectral softening 
from $\sim1$ to 2 or more during $\sim 200$ seconds \citep{kagawa2015}.
Thus the histogram of the obtained photon index of the extended
emission may be biased to soften.

If we consider the case that the energy spectrum of the extended
emission has the peak energy of $>10$ keV, its bolometric energy
should be modified.
The photon index in 2 -- 10 keV at the observed early phase is
$\sim1$ as shown in Figure \ref{fig:spectra} (e.g., GRB~050724, 060614,
070714B), and the one of the BAT spectrum before the XRT observations is
$\sim2$ (see Section \ref{subsec:batsp}).
Thus, the peak energy of the extended emission before its decaying phase
is thought to be around the lower threshold of the BAT energy range of
$\sim15$ keV.
Assuming the spectral shape of a broken power-law with the break energy
of 15 keV and low/high-energy photon indices of 1 and 2,
respectively, we estimate that the energy fluxes in 2 -- 150 keV is
larger by a factor of $\sim6$ than that in 2 -- 10 keV\footnote{The ratio
of the energy fluxes in 2 -- 150 keV to 2 -- 10 keV is $[(15-2)+15{\rm
log}(\frac{150}{15})]/(10-2)\sim6$.}. 
In such case, the ratio of $E_{\rm iso, EE}/E_{\rm iso, pro}$ becomes
closer to unity compared with the one of 2 -- 10 keV as shown in
Figure~\ref{fig:eiso}.
We conclude that
the majority of the extended emissions have the isotropic energy
comparable to or slightly less than that of the prompt emissions.

  \subsection{$L_{\rm EE} - \tau_{{\rm EE}}$ Correlation}
\label{subsec:l-tau}
Assuming that all of the selected SGRBs are followed by the exponential
decay, we show a scatter plot on $L_{{\rm EE}}$ - $\tau_{{\rm EE}}$ plane
(listed in Table \ref{table:results}) in Figure \ref{fig:ltau}.
There is a strong negative correlation between $L_{{\rm EE}}$ and
$\tau_{{\rm EE}}$ 
with the Spearman's rank order correlation coefficient of $-0.78$ and
chance probability of $8.2 \times 10^{-6}$.
By performing a power-law fit for the data, we obtain
\begin{eqnarray}
L_{{\rm EE}}(\tau_{{\rm EE}})=1.6^{+0.2}_{-0.2} \times 10^{50}\times
 \left( \frac{\tau_{{\rm EE}}}{{20\ {\rm
  s}}}\right)^{-3.3^{+0.1}_{-0.1}}\ {\rm erg~s}^{-1}, 
\end{eqnarray}
where $L_{{\rm EE}}(\tau_{{\rm EE}})$ is normalized at $\tau_{{\rm
EE}}=20$ seconds which is an observed shortest value in the fitting
results as listed in Table~\ref{table:results}.

Figure \ref{fig:ltau} shows lack of events on the upper-right (long and
bright) and lower-left (short and dim) areas. Such deserts could be caused
by observation bias which makes the apparent correlation.
Thus, we carefully consider the observation bias of the {\it Swift}/XRT
observation.
First, in the case of the long and bright events, if there existed such
events, the events should have been confidently detected.
However, no detection of such events indicates that intrinsically there
is no long and bright event.
Next, in the case of the short and dim events, such events might belong
to sub- or under-threshold events observed with the XRT and we
detailedly consider the XRT sensitivity of observation for the extended
emission.    

We consider that the XRT starts to observe a burst with a sensitivity,
$F_{{\rm XRT}} = 2 \times 10^{-10} \times (T_{{\rm XRT}}/1 {\rm
s})^{-1}$ erg~cm$^{-2}$~s$^{-1}$ (in 0.3 -- 10 keV) \citep{burrows2005},
at $T_{{\rm start}}$ second in the observer frame after a burst trigger,
where $T_{{\rm XRT}}$ is an integration time after the XRT starts an
observation.
The energy sensitivity of the XRT in a rest-frame energy band of $2-10$
keV and a rest-frame duration $\tau_{{\rm EE}}$ is described as
$4\pi d^2_LF_{{\rm XRT}} \tau_{{\rm EE}} P$ [erg],
where,
\begin{eqnarray}
 P = \frac{\int_{0.3~ {\rm keV}}^{10~{\rm keV}}N(E) \cdot
  EdE}{\int_{2/(1+z)~{\rm keV}}^{10/(1+z)~{\rm keV}}N(E) \cdot EdE},
\end{eqnarray}
and $d_L$ is a luminosity distance in unit of cm. Since the extended
emission at early phase has a hard energy spectrum as shown in Figure
\ref{fig:spectra} (e.g., GRB~050724, 060614, 070714B),
we make an assumption that the photon index $\Gamma$ for the
power-law spectrum of $N(E)$ equals to 1.
When, $T_{{\rm start}}$ seconds after a burst, the XRT observes the
extended emission for $\sim\tau_{{\rm EE}}$ seconds, the following
relation should be satisfied for the XRT to detect the extended emission,
\begin{eqnarray}
 \int^{T_{{\rm start}}/(1+z)+\tau_{\rm EE}}_{T_{{\rm start}}/(1+z)}L(t)dt >  4\pi
  d^2_L F_{{\rm XRT}} \tau_{{\rm EE}} P. 
\end{eqnarray}
Therefore, the observational luminosity limit of the XRT, $L_{{\rm limit,
XRT}}(\tau_{{\rm EE}})$, as a function of an e-folding time $\tau_{{\rm
EE}}$ with a parameter $z$ is provided with
\begin{eqnarray}
 L_{{\rm limit,XRT}}(\tau_{{\rm EE}}) = \frac{4\pi d^2_L F_{{\rm XRT}}
 \tau_{{\rm EE}} P}
 {\int^{T_{{\rm start}}/(1+z)+\tau_{\rm EE}}_{T_{{\rm start}}/(1+z)}{\rm
 exp}(-t/\tau_{{\rm EE}})dt}. 
\end{eqnarray}
In Figure \ref{fig:ltau}, we show the limits of $L_{{\rm limit,
XRT}}(\tau_{{\rm EE}})$, where $T_{{\rm start}}$ is provided with an
average value of 80 seconds.
Here, we adopt representative redshifts $z$ of 0.1 and 0.72,
which are similar to the nearest redshift of the observed {\it Swift}
SGRB ($z=0.111$ of GRB~061201) and the averaged redshift of SGRBs
observed by {\it Swift} with known redshift \citep{kisaka2017},
respectively.
Namely, the $L_{{\rm limit, XRT}}(\tau_{{\rm EE}})$ with $z=0.1$
and $0.72$ correspond to possibly the most optimistic and typical
luminosity limits, respectively.

Here, we consider lack of events with $\tau_{{\rm EE}} < 20$ s and
$L_{{\rm EE}} < 10^{49}$ erg~s$^{-1}$ (i.e., absence of dim GRBs with
short $\tau_{{\rm EE}}$) as illustrated in Figure \ref{fig:ltau}, which
mainly highlights the $L_{{\rm EE}}$ -- $\tau_{{\rm EE}}$ relation newly
found in this paper.
By considering that most of SGRBs occur at $z\sim0.72$ as the typical case,
a realistic luminosity limit should be represented with $z=0.72$.
The figure shows that the $L_{{\rm limit,XRT}}(\tau_{{\rm EE}})$ with
$z=0.72$ is dimmer by about one order of magnitude than the 
observed events with $\tau_{\rm EE} =$ 20 -- 30 s.
Thus we conclude that the events on such region is free from the
observation bias.

Then, we consider the nearby events with $z<0.72$, especially $z\sim0.1$, to
discuss such short $\tau_{{\rm EE}}$ events more.
For the $L_{{\rm limit,XRT}}(\tau_{{\rm EE}})$ with $z=0.1$, the
observations by the XRT are supposed to search for a large parameter
space of $L_{{\rm EE}} > L_{{\rm limit,XRT}}(\tau_{{\rm EE}})$ with $z =
0.1$. 
However, the $L_{{\rm limit ,XRT}}(\tau_{{\rm EE}})$ with $z=0.1$
is an optimistic case assuming the nearest {\it Swift} SGRB.
Since the number of SGRBs with $z\sim0.1$ observed with the XRT in this
plot is statistically limited at this moment, this luminosity limit
might be too optimistic. 
Therefore, we cannot fully reject a possibility that the $L_{\rm
EE}-\tau_{\rm EE}$ correlation for dimmer events with $L_{\rm
EE}<10^{47}$ erg~s$^{-1}$ are affected by the observation bias.
However, as described before, the correlation for brighter events is
the intrinsic property of the extended emission.

Finally, we conclude that there is the strong anti correlation between
$L_{{\rm EE}}$ and $\tau_{{\rm EE}}$ whose power-law index is about $-3.3$
although it is difficult to discuss the observation bias for the dimmer
events close to the luminosity limit.
This value is similar to an index of the luminosity -- duration plot in
\cite{kisaka2017}.
On the other hand, there are some works for luminosity -- time
correlation for LGRBs. 
\cite{willingale2010} shows a correlation between the peak luminosity
and peaking time of each pulse in 11 LGRBs and GRB050724 (considered to
be a SGRB candidate) with an index of $-2.0$.
\cite{dainotti2015} also indicates a correlation between the plateau's
luminosity at the end of the plateau phase and the duration of the
plateau emission with an index of $\sim -0.90$ in LGRBs.
These previous results suggest that the extended emission has unique
properties which is different from such long activities in LGRBs.
The steep index of the extended emission would be a key to revealing the
mechanism of the extended emission and even SGRBs.

  \subsection{Physical Exponential Decay Model of Extended Emission}
  \label{subsec:expmodel}
Several models suggest the exponentially decaying light curves. Here we
adopt the basic picture that a SGRB and following emissions originate from
a relativistic outflow launched from a merger remnant (a black hole or a
neutron star) of a binary neutron stars or a black hole - neutron star
binary. First, since the luminosity of Blandford-Znajek jet is
proportional to the square of the magnetic flux on the black hole,
$L \propto B^2$, the luminosity could decay due to decrease of the
magnetic flux \citep{bz1977}. Then, if the magnetic field energy
exponentially decays, the luminosity does, too. Such a magnetic energy
dissipation would be expected from fallback of the merger
ejecta \citep{kisaka2015}. The fallback matter drags the magnetic field
lines to the black hole because of the frozen-in condition, and
eventually forces the anti-directed magnetic fields to reconnect
\cite[see Figure 1 in][]{kisaka2015}. In this case, the duration of the
extended emission would be determined by the magnetic field dissipation,
not by the escape of the field line from the black hole considered in
the decay of the plateau emission phase \cite[$L\propto
t^{-40/9}$;][]{kisaka2015,kisaka2017}. It is noted that the energy
released due to magnetic reconnection is negligible for the energy
extracted by the Blandford-Znajek process \citep{kisaka2015}.

Second possibility is that the rotation energy loss rate of a star with
a split-monopole configuration follows the exponential decay after the
spin-down timescale, $E_{\rm rot}/L$, where $E_{\rm rot}$ is the
rotation energy of the remnant. The split monopole-like configuration
has been considered in the Blandford-Znajek jet model \citep{bz1977}.
If the duration of the extended emission is comparable to the spin-down
timescale, the following exponential decay of the luminosity is
expected. 
In this case, a relatively small value of the spin parameter of the black
hole is required \citep{nathanail2015} unless most of the rotation
energy is rapidly radiated by the gravitational wave. The total
radiation energy of the extended emission is $\sim10^{48}$ -- $10^{51}$ erg
(see Figure \ref{fig:eiso} or Table \ref{table:prompt}).
Assuming the radiation efficiency of $\sim0.1$
\cite[e.g.,][]{zhang2007}, the total energy is $\sim10^{49}$ -- $10^{52}$
erg, which corresponds to the rotation energy of $\sim3M_{\odot}$ black
hole with the dimensionless spin parameter of $a_{\ast}\sim$ 0.003 -- 0.1.
On the other hand, the dimensionless spin parameter of the collapsed BH
is $a_{\ast}\sim0.7$ from the numerical simulations of binary neutron
star merger \cite[e.g.,][]{shibata2006}.  

The split monopole-like configuration is also expected in the neutron
star engine case. Even if the neutron star has the dipole magnetic
field, the closed field lines could be truncated by the disk within the
light cylinder. The inner disk radius is determined by the pressure
balance between the magnetic field of the star and the accreting matter
\cite[e.g.,][]{ghosh1979}. If the inner disk radius is larger than the
co-rotation radius, where the Keplerian velocity equals to the
co-rotation velocity of the neutron star, the accreting matter gets the
angular momentum and the resultant wind makes the field lines open
\cite[propeller regime; e.g.,][]{illarionov1975}. If the inner radius of
the disk is steady in the propeller regime, the wind power follows the
exponential decay after the spin-down timescale
\cite[e.g.,][]{metzger2018}. Then, the mass accretion rate should be
constant, or be sufficiently high to keep the inner radius coincident
with the neutron star radius. Other possibility for the exponentially
decaying wind power is given by the exponentially decaying mass
accretion rate in the propeller regime \citep{gompertz2014}.  

The properties of additional long-lasting activities would allow us to
distinguish the models. A significant fraction of short GRBs shows a
long-lasting plateau emission with the luminosity of
$\sim 10^{46}$ -- $10^{47}$ erg~s$^{-1}$ and duration $\sim 10^3$ -- $10^4$
seconds in their light curves \citep{gompertz2013,rowlinson2013,lu2015}.
The fluences of extended and plateau emissions are of roughly
the same order of magnitude \citep{kisaka2017}.
The plateau emission is considered to be produced by an activity of the
central engine \citep{gompertz2014,kisaka2015}.
If the duration of the extended emission is determined by the spin-down
timescale, it is difficult to explain the plateau component whose energy
is comparable to or higher than the extended one.
The detailed systematic studies for the light curve shape and the energy
spectral distribution of the plateau emission will help to precisely
estimate the radiation energy, and separate the central engine
activities from the afterglow emission.

\section{Conclusion}
We find the following properties of the temporally extended X-ray
emission of SGRBs observed with the {\it Swift}/BAT and XRT.

\begin{enumerate}
 \item The light curves of the extended emissions following 23 of the 24
       ($\sim 96\%$) selected SGRBs with known redshifts are able to be
       described with the exponential temporal decay function
       with a rest-frame e-folding time of 20 -- 200 seconds.  
       For the power-law model, on the other hand, it is difficult to
       comprehensively describe the temporal behavior.
 \item The isotropic energy of the extended emission in 2 -- 10 keV
       calculated precisely by adopting the exponential decay is smaller
       by 0 -- 3 orders of magnitude than that of the prompt emission estimated
       bolometrically. 
 \item Between $L_{{\rm EE}}$ and $\tau_{\rm EE}$ there is a strong
       anti correlation with a steep power-law index of $\sim-3.3$.
\end{enumerate}    

To discuss the population of the extended emission and the observation
bias of the XRT in more detail, it is necessary to observe more SGRBs
with a dim extended emission by the {\it Swift}/XRT.
In 2020s, brand new observatories whose telescope employs a lobster-eye
optics covering sub- to several-keV energy band and a wide field of view,
such as Einstein Probe \citep{yuan2018}, ISS-TAO \citep{yacobi2018}, and
HiZ-GUNDAM (in prep.) will be launched, and then they can observe
the extended emissions from the brightening phase which can not be
observed by the XRT.
Furthermore, in third generation of GW observatories such as Einstein
Telescope \citep{sathyaprakash2012}, the detection alert of a GW signal
originate from a binary neutron stars can be sent 1 -- 20 hours before
the binary merges \citep{chan2018}.
Thus, future observatories with X-ray telescope(s) would observe
extended emissions ever since before the coalescence of the binary
neutron stars occurs if SGRBs and extended emissions originate from the
binary merger. 
Such observations with a better sensitivity than current detectors may
be unbiased ones.

\acknowledgments
This work was supported by
Grants-in-Aid for JSPS Research Fellow Grant Numbers JP18J13042 (YK),
JP16J06773 (SK), 
KAKENHI Grant Numbers JP16H06342 (DY), JP17H06362 (MA), 18H01245, JP18H01246  (SK),
JP18H01232 (RY),  
MEXT KAKENHI Grant Number JP18H04580 (DY),
JSPS Leading Initiative for Excellent Young Researchers program (MA),
and Sakigake 2018 Project of Kanazawa University (DY, MA).

\software{XSPEC \citep[v12.10.0][]{arnaud1996}}

\appendix

  \begin{table}[h]
   \caption{
   Sample of SGRBs referred from Swift GRB table.
   \footnote{\url{https://swift.gsfc.nasa.gov/archive/grb_table/}}
   }
   \label{table:grblist}
   \begin{tabular}{llcc} \hline \hline
    ID & Redshift & XRT Time to & Reference of Redshift \\ 
       &          & First Observation(s) &  \\ \hline
    GRB~050724  & 0.258  & 74\footnote{\cite{covino2005}}&  \cite{prochaska2005}   \\
    GRB~051221A & 0.5465 & 88.00 &  \cite{berger2005}   \\
    GRB~060614  & 0.125  & 91.40 &  \cite{fugazza2006}   \\
    GRB~060801  & 1.131  & 63.01 &  \cite{cucchiara2006}   \\
    GRB~061006  & 0.4377 & 156.58&  \cite{berger2007}   \\
    GRB~061201  & 0.111  & 81.32 &  \cite{berger2006}   \\
    GRB~070714B & 0.923  & 61.37 &  \cite{graham2009}   \\
    GRB~070724A & 0.457  & 66.76 &  \cite{cucchiara2007}   \\
    GRB~070809  & 0.2187 & 70.78 &  \cite{perley2008}   \\
    GRB~071227  & 0.383  & 79.09 &  \cite{davanzo2007}   \\
    GRB~080123  & 0.495  & 101.81&  \cite{leibler2010}   \\
    GRB~080905A & 0.1218 & 130.38&  \cite{rowlinson2010}   \\
    GRB~090426  & 2.609  & 84.62 &  \cite{levesque2009}   \\
    GRB~090510  & 0.903  & 94.10 &  \cite{rau2009}   \\
    GRB~100117A & 0.92   & 80.1  &  \cite{fong2011}   \\
    GRB~100625A & 0.453  & 48.26 &  \cite{fong2013}   \\
    GRB~100816A & 0.8049 & 87.31 &  \cite{gorosabel2010}   \\
    GRB~101219A & 0.718  & 221.92&  \cite{chornock2011}   \\
    GRB~111117A & 2.211  & 76.8\footnote{\cite{mangano2011}} &  \cite{selsing2018}   \\
    GRB~130603B & 0.3586 & 59.05 &  \cite{thone2013}   \\
    GRB~140903A & 0.351  & 59    &  \cite{troja2016}   \\
    GRB~150423A & 1.394  & 70.12 &  \cite{malesani2015}   \\
    GRB~150424A & 0.3    & 87.87 &  \cite{castro2015}   \\
    GRB~160410A & 1.717  & 82.89 &  \cite{selsing2016}   \\
    GRB~160624A & 0.483  & 73.72 &  \cite{cucchiara2016}   \\
    GRB~160821B & 0.16   & 65.97 &  \cite{levan2016}   \\
        \hline

   \end{tabular}

  \end{table}
 
\clearpage

\begin{longtable}{lc|cccc|l}
\caption{Light curve fitting results.}
\label{table:results}
\\
\hline

 \colhead{GRB} & \colhead{model} & \colhead{$L_{{\rm EE}}$} &
 \colhead{$\tau_{{\rm EE}}$}  & \colhead{$T_{{\rm EE}}$} &
 \colhead{$\alpha_{{\rm EE}}$} & \colhead{$\chi^2_{\nu}$(dof)}\\ 
\hline \hline
\endfirsthead
\multicolumn{7}{c}%
{\tablename\ \thetable\ -- \textit{Continued from previous page}} \\
 \hline
  \colhead{GRB} & \colhead{model} & \colhead{$L_{{\rm EE}}$} &
 \colhead{$\tau_{{\rm EE}}$}  & \colhead{$T_{{\rm EE}}$} &
 \colhead{$\alpha_{{\rm EE}}$} & \colhead{$\chi^2_{\nu}$(dof)}\\
 \hline
\hline
\endhead
\hline \multicolumn{7}{r}{\textit{Continued on next page}} \\ 
\endfoot
\hline
\endlastfoot
 \hline

 050724 & EXP & (7.04 $\pm$ 0.36) $\times$ $10^ { 48 } $  & 46.3 $\pm$ 1.1 & & & 4.80 (61) \\
& PL(BH)  & (2.48 $\pm$ 0.67) $\times$ $10^ { 49 } $ & & (8.34 $\pm$ 0.94) $\times$ $10^ { 1 } $ & 4.44(fix) & 6.37(62) \\
& PL(MG) & --- & & --- & 2(fix) & 14.39(62) \\
\hline
051221A & EXP & (1.50 $\pm$ 0.22) $\times$ $10^ { 47 } $  & 209.1 $\pm$ 19.6 & & & 2.33 (10) \\
& PL(BH)  & (2.17 $\pm$ 0.51) $\times$ $10^ { 47 } $ & & (4.72 $\pm$ 0.73) $\times$ $10^ { 2 } $ & 4.44(fix) & 1.36(11) \\
& PL(MG) & (7.72 $\pm$ 5.10) $\times$ $10^ { 47 } $ & & (5.08 $\pm$ 2.10) $\times$ $10^ { 1 } $ & 2(fix) & 1.12(11) \\
\hline
060614 & EXP & (1.79 $\pm$ 0.06) $\times$ $10^ { 49 } $  & 40.3 $\pm$ 0.4 & & & 3.30 (100) \\
& PL(BH)  & (9.85 $\pm$ 0.54) $\times$ $10^ { 50 } $ & & (3.06 $\pm$ 0.05) $\times$ $10^ { 1 } $ & 4.44(fix) & 6.26(101) \\
& PL(MG) & --- & & --- & 2(fix) & 48.27(101) \\
\hline
060801 & EXP & (4.11 $\pm$ 0.45) $\times$ $10^ { 47 } $  & 99.4 $\pm$ 10.2 & & & 0.70 (15) \\
& PL(BH)  & (4.58 $\pm$ 0.62) $\times$ $10^ { 47 } $ & & (3.41 $\pm$ 0.45) $\times$ $10^ { 2 } $ & 4.44(fix) & 0.95(16) \\
& PL(MG) & (5.41 $\pm$ 0.83) $\times$ $10^ { 47 } $ & & (1.05 $\pm$ 0.17) $\times$ $10^ { 2 } $ & 2(fix) & 1.02(16) \\
\hline
061006 & EXP & (1.21 $\pm$ 0.58) $\times$ $10^ { 46 } $  & 184.3 $\pm$ 70.9 & & & 0.82 (3) \\
& PL(BH)  & (1.64 $\pm$ 1.01) $\times$ $10^ { 46 } $ & & (5.47 $\pm$ 2.35) $\times$ $10^ { 2 } $ & 4.44(fix) & 0.59(4) \\
& PL(MG) & (3.90 $\pm$ 3.89) $\times$ $10^ { 46 } $ & & (9.10 $\pm$ 6.60) $\times$ $10^ { 1 } $ & 2(fix) & 0.49(4) \\
\hline
061201 & EXP & (2.94 $\pm$ 2.20) $\times$ $10^ { 45 } $  & 86.1 $\pm$ 46.6 & & & 0.94 (19) \\
& PL(BH)  & (3.55 $\pm$ 2.84) $\times$ $10^ { 45 } $ & & (3.04 $\pm$ 2.15) $\times$ $10^ { 2 } $ & 4.44(fix) & 0.91(20) \\
& PL(MG) & (1.34 $\pm$ 5.72) $\times$ $10^ { 46 } $ & & (2.87 $\pm$ 7.92) $\times$ $10^ { 1 } $ & 2(fix) & 1.10(20) \\
\hline
070714B & EXP & (7.11 $\pm$ 2.11) $\times$ $10^ { 49 } $  & 18.3 $\pm$ 1.5 & & & 2.47 (45) \\
& PL(BH)  & (5.50 $\pm$ 1.28) $\times$ $10^ { 52 } $ & & (7.38 $\pm$ 0.45) $\times$ $10^ { 0 } $ & 4.44(fix) & 4.42(46) \\
& PL(MG) & --- & & --- & 2(fix) & 9.31(46) \\
\hline
070724A & EXP & (2.84 $\pm$ 0.80) $\times$ $10^ { 48 } $  & 32.2 $\pm$ 2.5 & & & 3.12 (12) \\
& PL(BH)  & (3.78 $\pm$ 3.13) $\times$ $10^ { 49 } $ & & (3.74 $\pm$ 0.91) $\times$ $10^ { 1 } $ & 4.44(fix) & 1.96(13) \\
& PL(MG) & (1.79 $\pm$ 1.63) $\times$ $10^ { 53 } $ & & (0.70 $\pm$ 0.33) $\times$ $10^ { -1 } $ & 2(fix) & 4.52(13) \\
\hline
070809 & EXP & (6.11 $\pm$ 3.24) $\times$ $10^ { 45 } $  & 78.2 $\pm$ 24.7 & & & 0.58 (15) \\
& PL(BH)  & (1.20 $\pm$ 1.12) $\times$ $10^ { 46 } $ & & (1.71 $\pm$ 0.93) $\times$ $10^ { 2 } $ & 4.44(fix) & 0.60(16) \\
& PL(MG) & (4.79 $\pm$ 2.25) $\times$ $10^ { 46 } $ & & (2.00 $\pm$ 2.47) $\times$ $10^ { 1 } $ & 2(fix) & 0.74(16) \\
\hline
071227 & EXP & (1.97 $\pm$ 0.55) $\times$ $10^ { 48 } $  & 36.7 $\pm$ 2.6 & & & 1.89 (7) \\
& PL(BH)  & (1.37 $\pm$ 1.29) $\times$ $10^ { 55 } $ & & (1.64 $\pm$ 0.35) $\times$ $10^ { 0 } $ & 4.44(fix) & 1.87(8) \\
& PL(MG) & --- & & --- & 2(fix) & 10.28(8) \\
\hline
080123 & EXP & (1.69 $\pm$ 0.41) $\times$ $10^ { 49 } $  & 26.7 $\pm$ 1.5 & & & 2.20 (10) \\
& PL(BH)  & (1.45 $\pm$ 0.94) $\times$ $10^ { 56 } $ & & (1.17 $\pm$ 0.17) $\times$ $10^ { 0 } $ & 4.44(fix) & 3.44(11) \\
& PL(MG) & --- & & --- & 2(fix) & 31.01(11) \\
\hline
080905A & EXP & (1.12 $\pm$ 0.27) $\times$ $10^ { 46 } $  & 104.0 $\pm$ 15.9 & & & 0.47 (5) \\
& PL(BH)  & (1.85 $\pm$ 0.16) $\times$ $10^ { 46 } $ & & (2.82 $\pm$ 0.14) $\times$ $10^ { 2 } $ & 4.44(fix) & 0.54(6) \\
& PL(MG) & (2.62 $\pm$ 0.24) $\times$ $10^ { 47 } $ & & (1.69 $\pm$ 0.09) $\times$ $10^ { 1 } $ & 2(fix) & 0.88(6) \\
\hline
090426 & EXP & (6.49 $\pm$ 0.65) $\times$ $10^ { 47 } $  & 178.4 $\pm$ 23.3 & & & 0.95 (20) \\
& PL(BH)  & (7.34 $\pm$ 0.79) $\times$ $10^ { 47 } $ & & (6.49 $\pm$ 1.11) $\times$ $10^ { 2 } $ & 4.44(fix) & 0.95(21) \\
& PL(MG) & (8.46 $\pm$ 1.03) $\times$ $10^ { 47 } $ & & (2.22 $\pm$ 0.33) $\times$ $10^ { 2 } $ & 2(fix) & 1.03(21) \\
\hline
090510 & EXP & $<3.93 \times10^ { 49 } $  & 20.3 $\pm$ 1.2 & & & 1.62 (33) \\
& PL(BH)  & (4.30 $\pm$ 8.00) $\times$ $10^ { 56 } $ & & (0.83 $\pm$ 0.36) $\times$ $10^ { 0 } $ & 4.44(fix) & 1.74(34) \\
& PL(MG) & (4.50 $\pm$ 11.80) $\times$ $10^ { 52 } $ & & (0.16 $\pm$ 0.22) $\times$ $10^ { 0 } $ & 2(fix) & 2.92(34) \\
\hline
100117A & EXP & (1.07 $\pm$ 0.43) $\times$ $10^ { 49 } $  & 32.5 $\pm$ 3.5 & & & 2.94 (9) \\
& PL(BH)  & (5.34 $\pm$ 4.64) $\times$ $10^ { 52 } $ & & (8.15 $\pm$ 1.72) $\times$ $10^ { 0 } $ & 4.44(fix) & 1.74(10) \\
& PL(MG) & --- & & --- & 2(fix) & 7.40(10) \\
\hline
100625A & EXP & (9.42 $\pm$ 2.14) $\times$ $10^ { 45 } $  & 144.0 $\pm$ 34.2 & & & 0.44 (6) \\
& PL(BH)  & (1.04 $\pm$ 0.27) $\times$ $10^ { 46 } $ & & (4.94 $\pm$ 1.41) $\times$ $10^ { 2 } $ & 4.44(fix) & 0.46(7) \\
& PL(MG) & (1.26 $\pm$ 0.45) $\times$ $10^ { 46 } $ & & (1.46 $\pm$ 0.57) $\times$ $10^ { 2 } $ & 2(fix) & 0.62(7) \\
\hline
100816A & EXP & $<2.96 \times 10^ { 49 } $  & 18.9 $\pm$ 0.8 & & & 1.30 (26) \\
& PL(BH)  & (4.40 $\pm$ 3.48) $\times$ $10^ { 48 } $ & & (1.17 $\pm$ 0.36) $\times$ $10^ { 2 } $ & 4.44(fix) & 3.66(27) \\
& PL(MG) & (4.24 $\pm$ 3.17) $\times$ $10^ { 52 } $ & & (0.23 $\pm$ 0.09) $\times$ $10^ { 0 } $ & 2(fix) & 2.57(27) \\
\hline
101219A & EXP & (4.09 $\pm$ 0.68) $\times$ $10^ { 47 } $  & 81.7 $\pm$ 16.8 & & & 3.57 (9) \\
& PL(BH)  & (5.07 $\pm$ 0.91) $\times$ $10^ { 47 } $ & & (2.58 $\pm$ 0.42) $\times$ $10^ { 2 } $ & 4.44(fix) & 3.54(10) \\
& PL(MG) & (8.80 $\pm$ 3.64) $\times$ $10^ { 47 } $ & & (5.50 $\pm$ 1.92) $\times$ $10^ { 1 } $ & 2(fix) & 4.26(10) \\
\hline
111117A & EXP & (1.61 $\pm$ 0.79) $\times$ $10^ { 48 } $  & 35.0 $\pm$ 12.6 & & & 1.01 (4) \\
& PL(BH)  & (2.47 $\pm$ 2.39) $\times$ $10^ { 48 } $ & & (9.54 $\pm$ 6.34) $\times$ $10^ { 1 } $ & 4.44(fix) & 1.02(5) \\
& PL(MG) & (5.89 $\pm$ 4.52) $\times$ $10^ { 48 } $ & & (1.75 $\pm$ 0.98) $\times$ $10^ { 1 } $ & 2(fix) & 1.22(5) \\
\hline
130603B & EXP & (5.99 $\pm$ 3.91) $\times$ $10^ { 46 } $  & 140.3 $\pm$ 102.8 & & & 1.18 (17) \\
& PL(BH)  & (5.62 $\pm$ 3.05) $\times$ $10^ { 46 } $ & & (6.75 $\pm$ 6.27) $\times$ $10^ { 2 } $ & 4.44(fix) & 1.19(18) \\
& PL(MG) & (1.29 $\pm$ 4.32) $\times$ $10^ { 47 } $ & & (7.93 $\pm$ 22.00) $\times$ $10^ { 1 } $ & 2(fix) & 1.23(18) \\
\hline
140903A & EXP & (2.53 $\pm$ 1.83) $\times$ $10^ { 46 } $  & 47.1 $\pm$ 18.4 & & & 2.02 (50) \\
& PL(BH)  & (7.97 $\pm$ 12.80) $\times$ $10^ { 46 } $ & & (8.71 $\pm$ 6.30) $\times$ $10^ { 1 } $ & 4.44(fix) & 2.01(51) \\
& PL(MG) & (4.16 $\pm$ 17.10) $\times$ $10^ { 51 } $ & & (0.75 $\pm$ 1.61) $\times$ $10^ { -1 } $ & 2(fix) & 2.30(51) \\
\hline
150423A & EXP & (8.23 $\pm$ 2.65) $\times$ $10^ { 46 } $  & 121.8 $\pm$ 48.4 & & & 0.84 (3) \\
& PL(BH)  & (9.83 $\pm$ 4.00) $\times$ $10^ { 46 } $ & & (3.78 $\pm$ 2.05) $\times$ $10^ { 2 } $ & 4.44(fix) & 0.83(4) \\
& PL(MG) & (1.28 $\pm$ 1.00) $\times$ $10^ { 47 } $ & & (1.06 $\pm$ 0.91) $\times$ $10^ { 2 } $ & 2(fix) & 1.11(4) \\
\hline
150424A & EXP & (2.37 $\pm$ 0.28) $\times$ $10^ { 48 } $  & 44.5 $\pm$ 2.0 & & & 2.54 (45) \\
& PL(BH)  & (2.59 $\pm$ 0.90) $\times$ $10^ { 49 } $ & & (5.33 $\pm$ 0.59) $\times$ $10^ { 1 } $ & 4.44(fix) & 4.12(46) \\
& PL(MG) & --- & & --- & 2(fix) & 12.12(46) \\
\hline
160410A & EXP & (5.85 $\pm$ 2.97) $\times$ $10^ { 49 } $  & 25.4 $\pm$ 8.7 & & & 1.01 (2) \\
& PL(BH)  & (1.44 $\pm$ 0.12) $\times$ $10^ { 50 } $ & & (5.44 $\pm$ 0.21) $\times$ $10^ { 1 } $ & 4.44(fix) & 0.84(3) \\
& PL(MG) & --- & & --- & 2(fix) & 48.20(3) \\
\hline
160624A & EXP & (9.89 $\pm$ 6.37) $\times$ $10^ { 47 } $  & 35.6 $\pm$ 7.6 & & & 5.71 (1) \\
& PL(BH)  & (3.70 $\pm$ 2.98) $\times$ $10^ { 48 } $ & & (6.42 $\pm$ 1.91) $\times$ $10^ { 1 } $ & 4.44(fix) & 2.47(2) \\
& PL(MG) & (1.95 $\pm$ 4.80) $\times$ $10^ { 52 } $ & & (0.17 $\pm$ 0.22) $\times$ $10^ { 0 } $ & 2(fix) & 2.82(2) \\
\hline
160821B & EXP & (2.75 $\pm$ 0.35) $\times$ $10^ { 47 } $  & 49.8 $\pm$ 2.3 & & & 2.63 (16) \\
& PL(BH)  & (2.66 $\pm$ 0.83) $\times$ $10^ { 48 } $ & & (6.11 $\pm$ 0.62) $\times$ $10^ { 1 } $ & 4.44(fix) & 4.35(17) \\
 & PL(MG) & --- & & --- & 2(fix) & 12.05(17) \\

\end{longtable}

\clearpage

    \begin{table}[h]
     \caption{
     $E_{{\rm iso, EE}}$, $E_{{\rm iso, pro}}$ and $E_{{\rm peak}}$ of the
     prompt emission of the selected SGRBs detected with WIND/Konus,
     {\it Fermi}/GBM, or {\it Suzaku}/WAM coincided with the {\it
     Swift}/BAT.\footnote{The errors on $E_{{\rm iso, pro}}$ and  
   $E_{{\rm peak}}$ estimated with the WIND/Konus or {\it Suzaku}/WAM data
   correspond to 90\% confidence level and with the {\it Fermi}/GBM data
     correspond to 1$\sigma$ (68\%) confidence level.}
   }
   \label{table:prompt}
   \begin{tabular}{lcclccc} \hline \hline
    ID          &$E_{{\rm iso,EE}}$& $E_{{\rm iso,pro}}$ &$E_{{\rm
                peak}}$\footnote{The observer-frame energy}&Prompt
		    Detector& Energy Range  & Reference \\ 
                &[erg]&     [erg]      &[keV]     && for Prompt [keV]         &  \\ \hline
    GRB~051221A &$(3.13 \pm 0.55) \times 10^{49}$
  	        &$2.39^{+0.07}_{-1.27} \times 10^{51}$&$402_{-72}^{+93}$&Konus&$20-2000$
  		&\cite{golenetskii2005} \\ 
    GRB~060614  &$(7.21 \pm 0.25) \times10^ { 50}$
  	        & $2.90^{+0.20}_{-0.89} \times 10^{50}$&$302_{-85}^{+214}$&Konus&$20-2000$
  		&\cite{golenetskii2006a}\\
    GRB~061006  &$(2.23 \pm 1.25) \times10^ { 48}$
  	        & $1.69^{+0.15}_{-0.91} \times 10^{51}$&$664_{-144}^{+227}$&Konus&$20-2000$
  		&\cite{golenetskii2006b}\\
    GRB~061201  &$(2.53 \pm 2.34) \times10^ { 47}$
  	        & $1.48^{+0.19}_{-1.23} \times 10^{50}$&$873_{-284}^{+458}$&Konus&$20-3000$
  		&\cite{golenetskii2006c}\\
    GRB~070714B &$(1.30 \pm 0.40) \times10^ { 51}$
  	        & $8.01^{+2.82}_{-1.30} \times 10^{51}$&$1120_{-380}^{+780}$&WAM  &$15-2000$
  		&\cite{ohno2007}   \\
    GRB~071227  &$(7.23 \pm 2.02) \times10^ { 49}$
  	        & $5.73^{+0.72}_{-0.72} \times 10^{50}$&$\sim1000$ &Konus&$20-1300$
  		&\cite{golenetskii2007} \\
    GRB~100117A &$(3.48 \pm 1.46) \times10^ { 50}$
         	& $8.82^{+1.08}_{-1.08} \times 10^{50}$&$287^{+74}_{-50}$ &GBM  &$8 -1000$
  	        &\cite{paciesas2010}   \\
    GRB~100625A &$(1.36 \pm 0.45) \times10^ { 48}$
         	& $6.69^{+0.25}_{-0.25} \times 10^{50}$&$509^{+78}_{-62}$&GBM  &$8 -1000$
  	        &\cite{bhat2010} \\
    GRB~101219A &$(3.34 \pm 0.88) \times10^ { 49}$
        	& $4.70^{+0.65}_{-0.65} \times 10^{51}$&$490_{-79}^{+103}$&Konus&$20-10000$
  	        &\cite{golenetskii2010} \\
    GRB~111117A &$(5.63 \pm 3.44) \times10^ { 49}$
                & $7.48^{+0.22}_{-0.22} \times 10^{51}$&$\sim370$&GBM  &$10-1000$
  	        &\cite{foley2011}  \\
    GRB~130603B &$(8.39 \pm 7.84) \times10^ { 48}$
        	& $2.06^{+0.22}_{-0.22} \times 10^{51}$&$660\pm100$&Konus&$20-10000$
  	        &\cite{golenetskii2013} \\
    GRB~150424A &$(1.05 \pm 0.13) \times10^ { 50}$
        	& $3.90^{+0.24}_{-0.24} \times 10^{51}$&$919_{-76}^{+82}$&Konus&$20-10000$
  	        &\cite{golenetskii2015} \\
    GRB~160410A &$(1.49 \pm 0.91) \times10^ { 51}$
        	& $8.53^{+2.13}_{-2.13} \times 10^{52}$&$1416_{-356}^{+528}$&Konus&$20-10000$
  	        &\cite{frederiks2016}   \\
    GRB~160624A &$(3.52 \pm 2.39) \times10^ { 49}$
  	        &$3.01^{+0.29}_{-0.29} \times 10^{50}$&$841\pm358$&GBM
                &$10-1000$&\cite{hamburg2016}  \\ 
    GRB~160821B &$(1.37 \pm 0.18) \times10^ { 49}$
  	        &$9.86^{+1.12}_{-1.12} \times 10^{49}$&$84\pm19$&GBM
                &$10-1000$&\cite{stanbro2016} \\ 
    \hline

   \end{tabular}
    \end{table}

 \begin{figure}
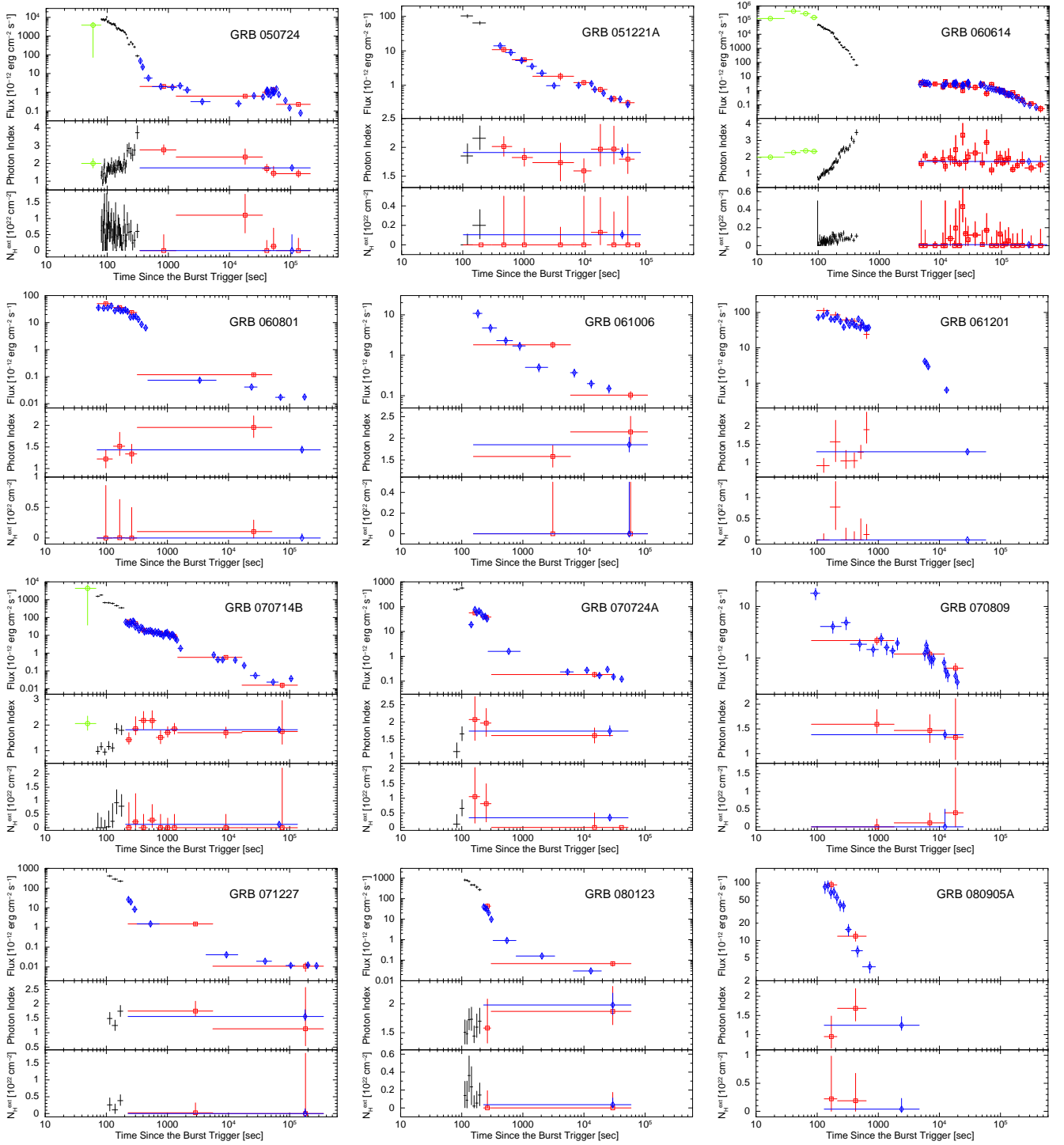

  \begin{center}
   
   \includegraphics[angle=270,scale=0.23]{050724.ps}
   \includegraphics[angle=270,scale=0.23]{051221A.ps}
   \includegraphics[angle=270,scale=0.23]{060614.ps}
   \includegraphics[angle=270,scale=0.23]{060801.ps}
   \includegraphics[angle=270,scale=0.23]{061006.ps}
   \includegraphics[angle=270,scale=0.23]{061201.ps}
   \includegraphics[angle=270,scale=0.23]{070714B.ps}
   \includegraphics[angle=270,scale=0.23]{070724A.ps}
   \includegraphics[angle=270,scale=0.23]{070809.ps}
   \includegraphics[angle=270,scale=0.23]{071227.ps}
   \includegraphics[angle=270,scale=0.23]{080123.ps}
   \includegraphics[angle=270,scale=0.23]{080905A.ps}
  \end{center}
  \caption{
  The results of power-law fitting for the spectra in each time bin
  intervals. (Top panels) the energy flux in units of
  $10^{-12}~$erg~cm$^{-2}$~s$^{-1}$ in the energy band of 2/(1+$z$) to  
  10/(1+$z$) (corresponding to the $2-10$ keV energy band in each rest
  frame). (Middle panels) the photon index $\Gamma$. (Bottom panels) the
  extra-galactic column density $N_{{\rm H}}^{{\rm ext}}$(see Section
  \ref{subsec:xrtsp} for the XRT data and Section \ref{subsec:batsp} for
  the BAT data). The green-opened circles show the single power-law
  fitting results for the BAT data. The black crosses and red squares
  corresponds to the results of the time-resolved analysis of the XRT's
  WT and PC mode data, respectively. The blue diamonds show the results
  of the time-averaged analysis for the data of PC mode of the XRT.  
  }
  \label{fig:spectra}
 \end{figure}
 
 \addtocounter{figure}{-1}
 \begin{figure}[tbp]
  \begin{center}
   \includegraphics[angle=270,scale=0.23]{090426.ps}
   \includegraphics[angle=270,scale=0.23]{090510.ps}
   \includegraphics[angle=270,scale=0.23]{100117A.ps}
   \includegraphics[angle=270,scale=0.23]{100625A.ps}
   \includegraphics[angle=270,scale=0.23]{100816A.ps}
   \includegraphics[angle=270,scale=0.23]{101219A.ps}
   \includegraphics[angle=270,scale=0.23]{111117A.ps}
   \includegraphics[angle=270,scale=0.23]{130603B.ps}
   \includegraphics[angle=270,scale=0.23]{140903A.ps}
   \includegraphics[angle=270,scale=0.23]{150423A.ps}
   \includegraphics[angle=270,scale=0.23]{150424A.ps}
   \includegraphics[angle=270,scale=0.23]{160410A.ps}
   \includegraphics[angle=270,scale=0.23]{160624A.ps}
   \includegraphics[angle=270,scale=0.23]{160821B.ps}
   \caption{
   Continued.
   }
   \label{fig:spectra}
  \end{center}
 \end{figure}


 \begin{figure}[tbp]
  \begin{center}
   \includegraphics[angle=0,scale=0.28]{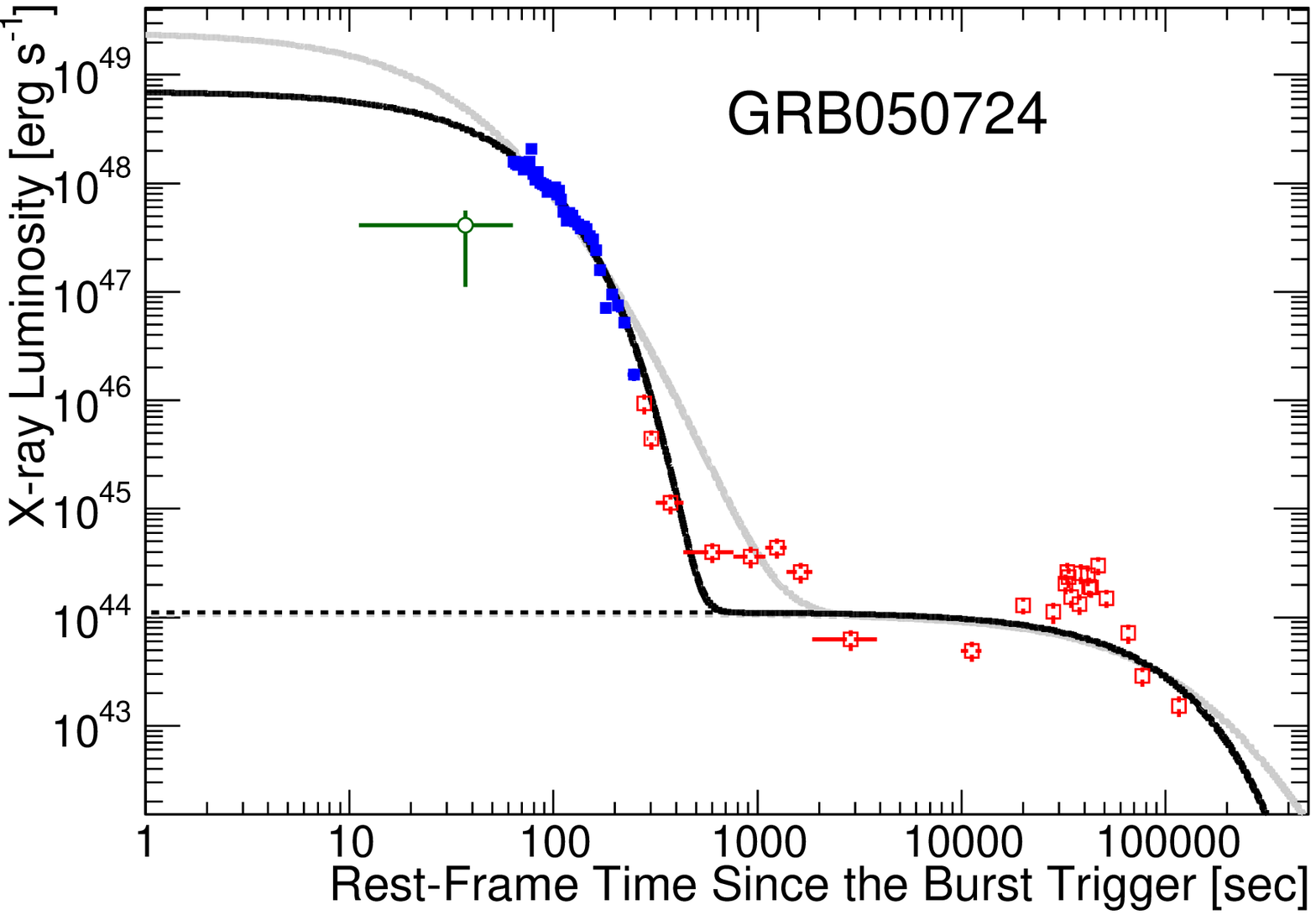}
   \includegraphics[angle=0,scale=0.28]{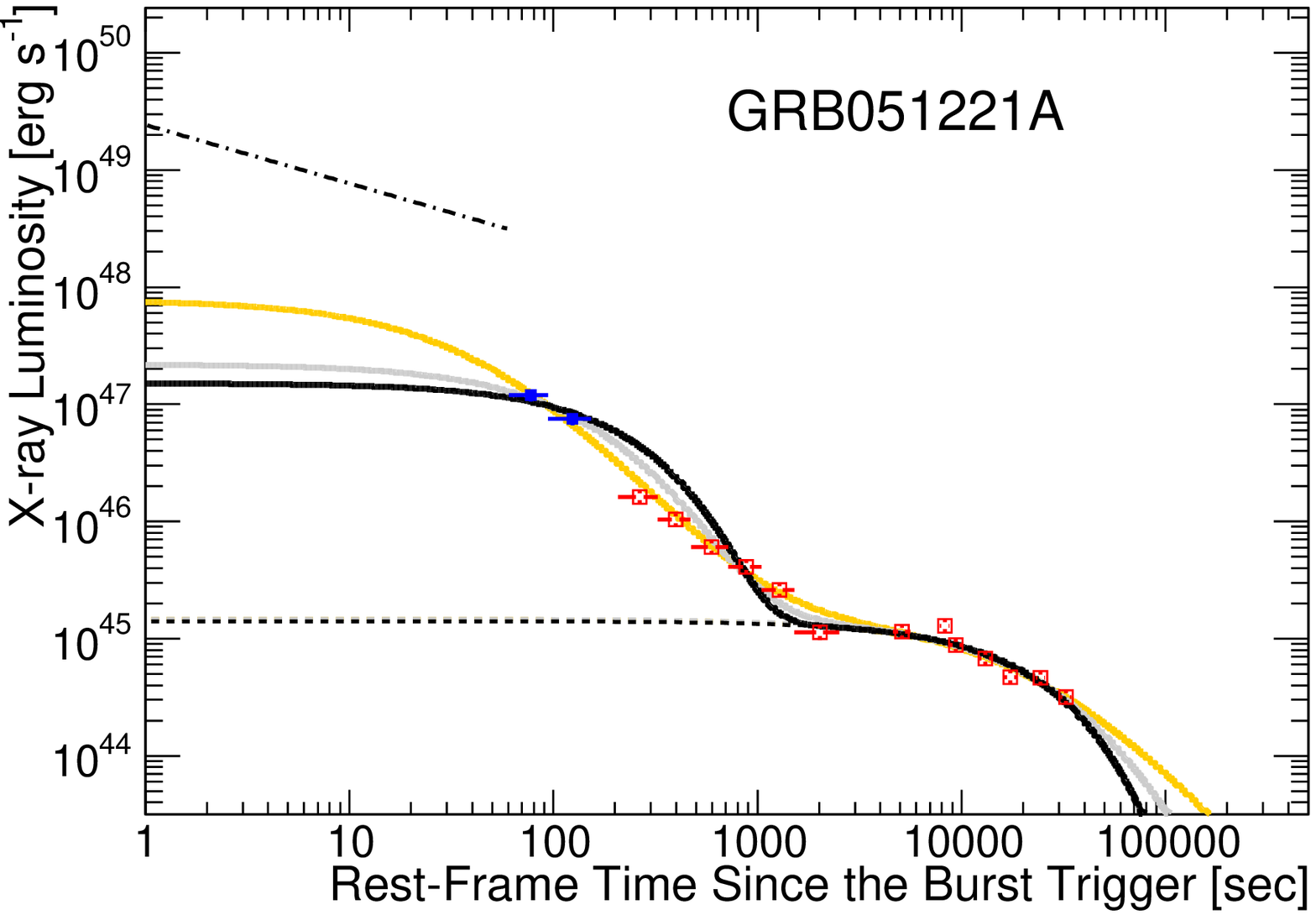}
   \includegraphics[angle=0,scale=0.28]{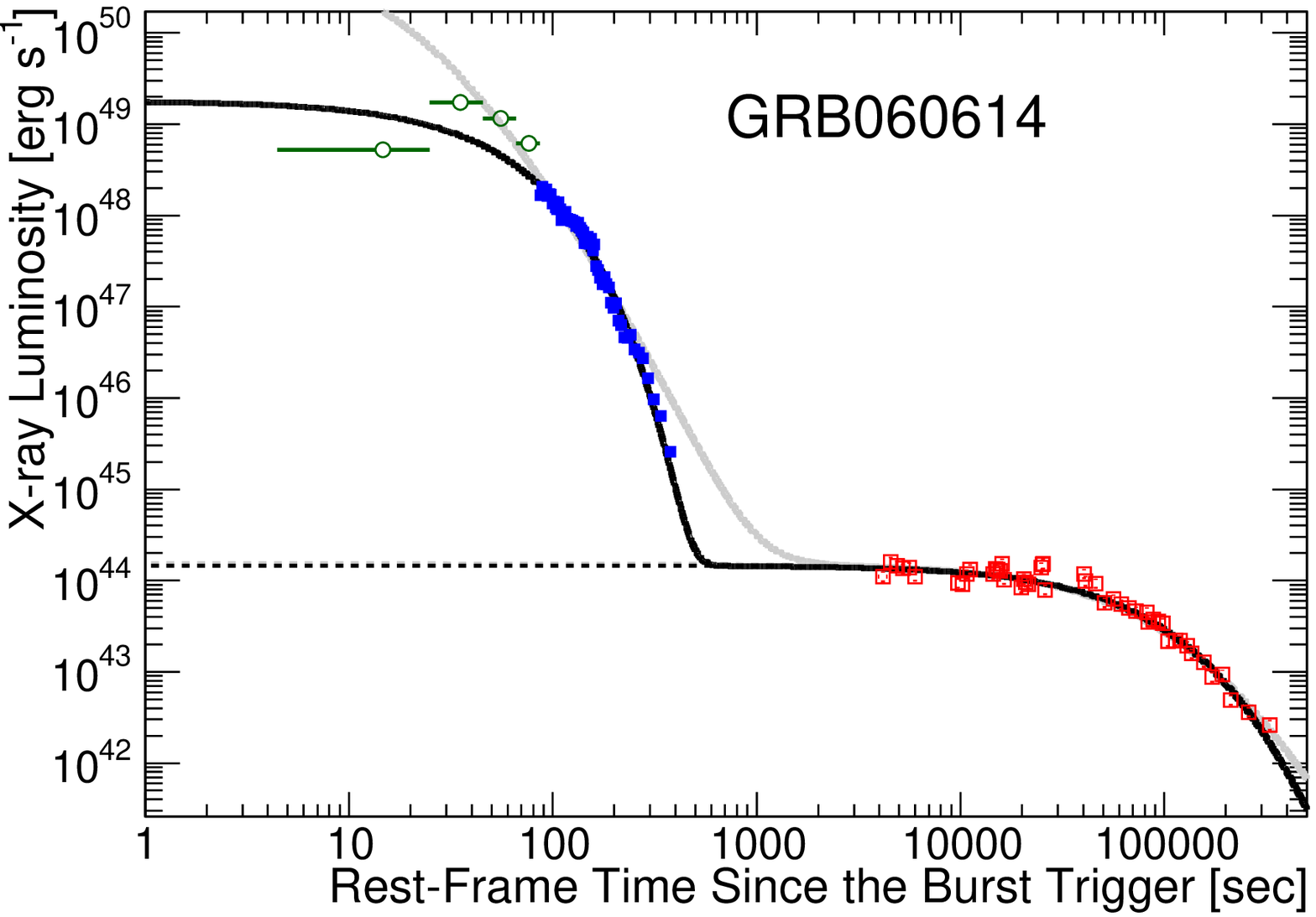}
   \includegraphics[angle=0,scale=0.28]{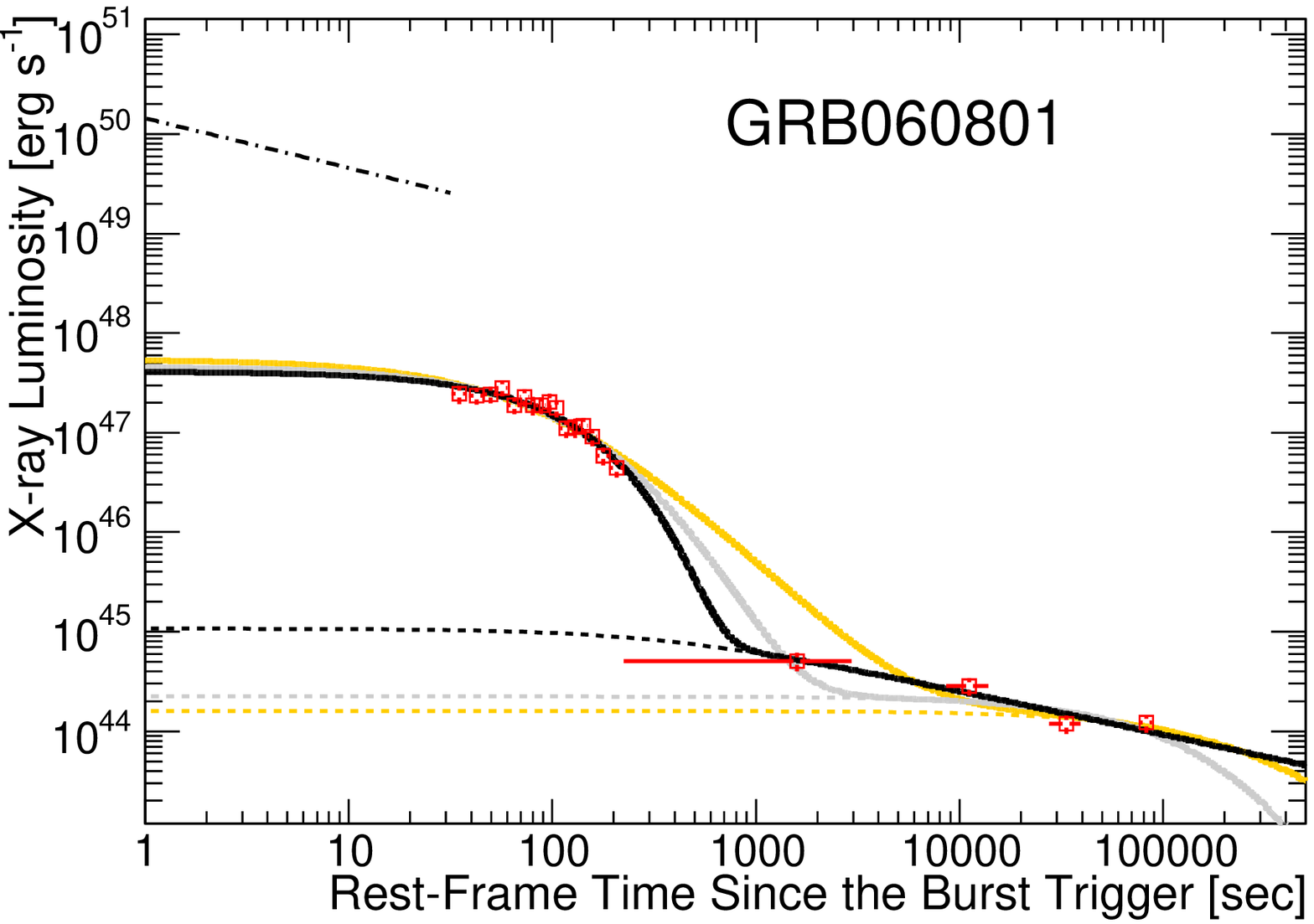}
   \includegraphics[angle=0,scale=0.28]{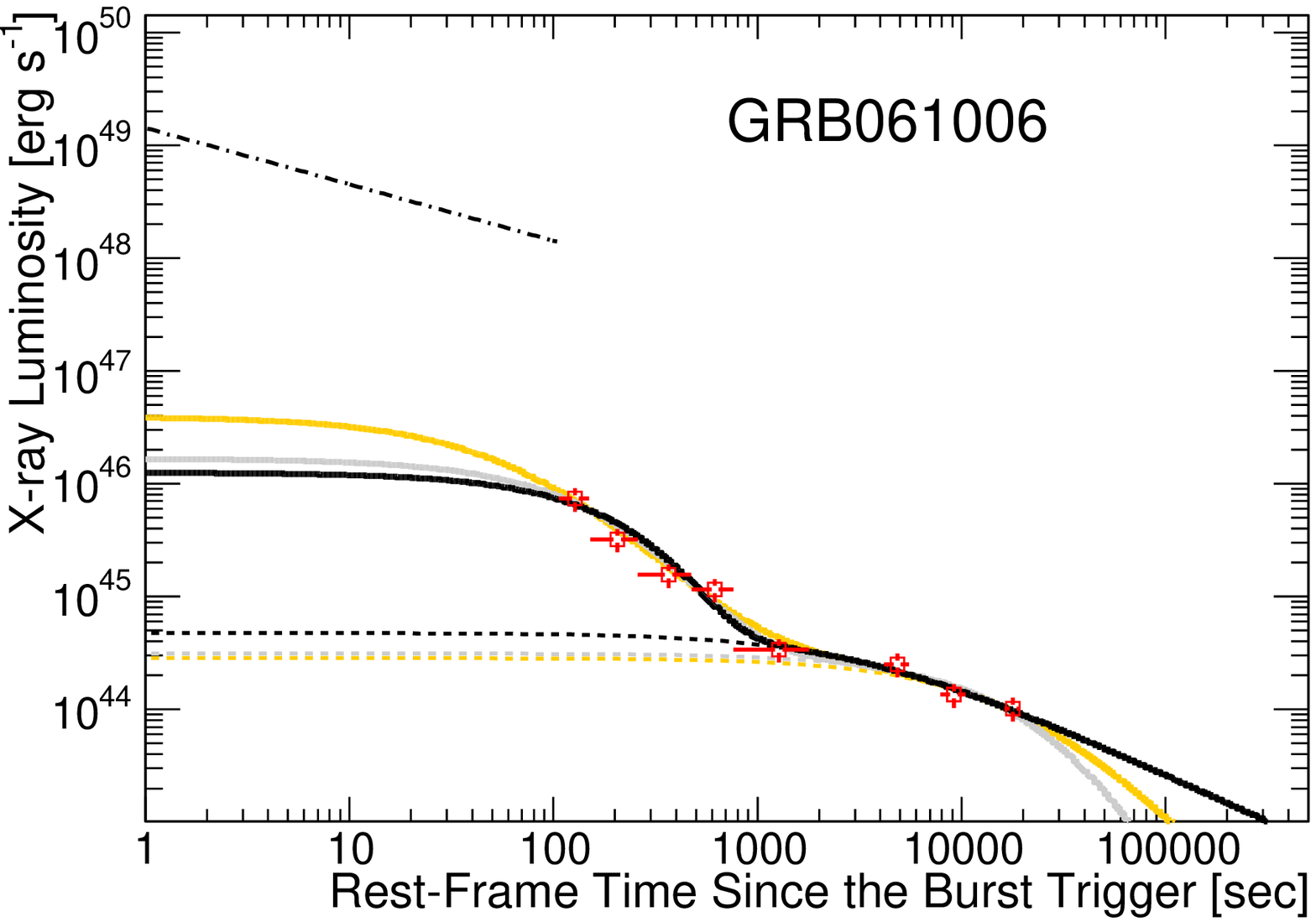}
   \includegraphics[angle=0,scale=0.28]{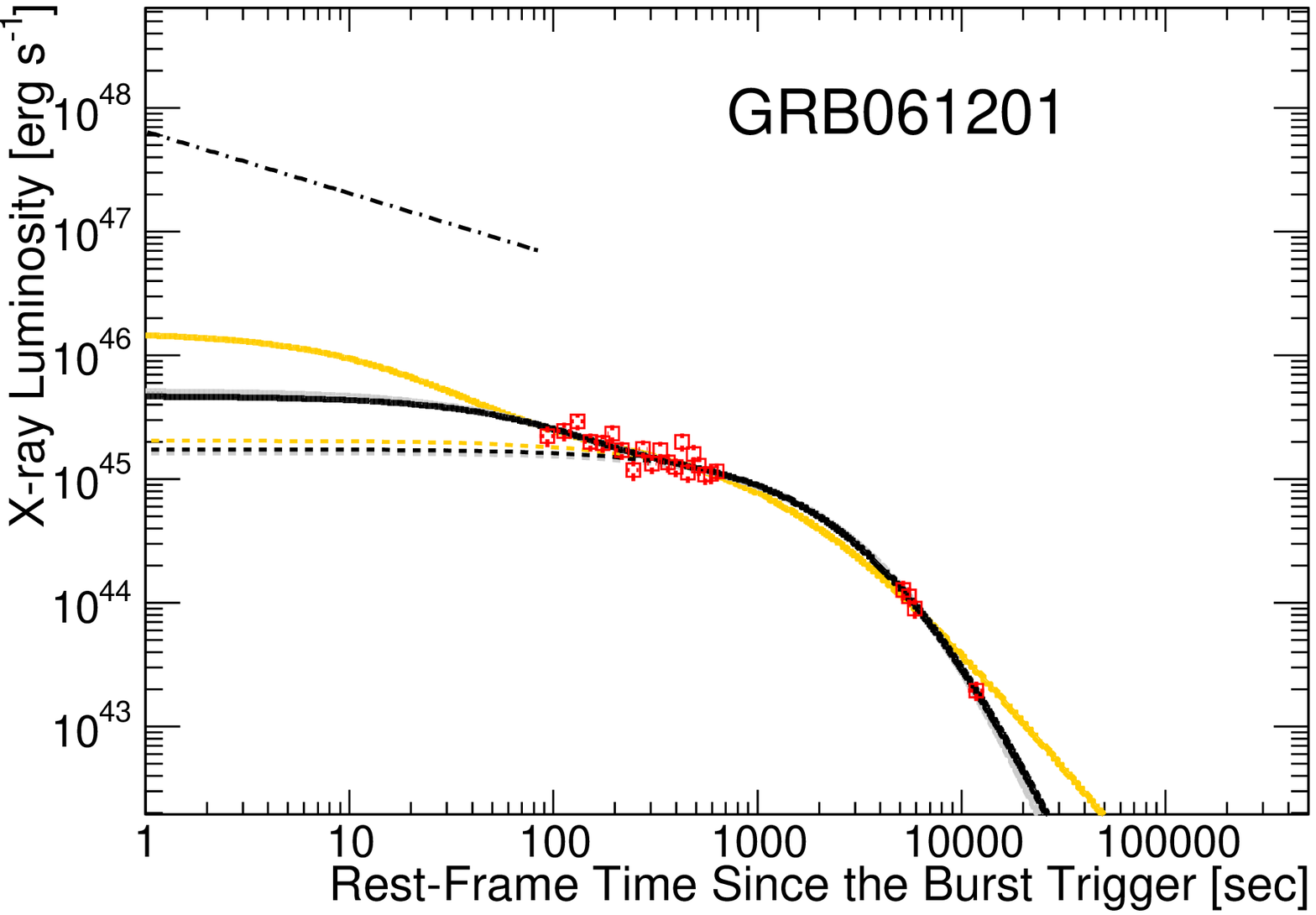}
   \includegraphics[angle=0,scale=0.28]{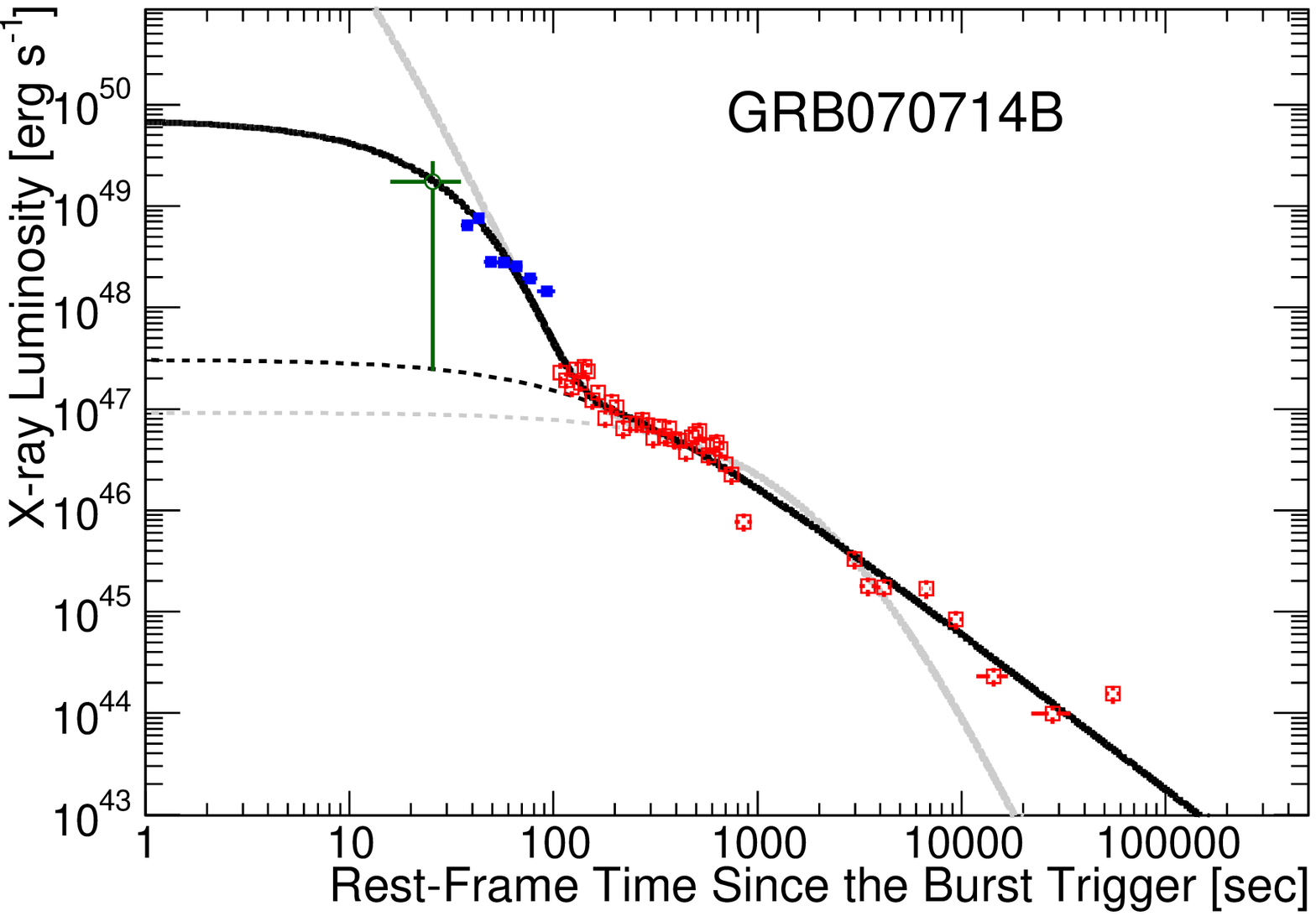}
   \includegraphics[angle=0,scale=0.28]{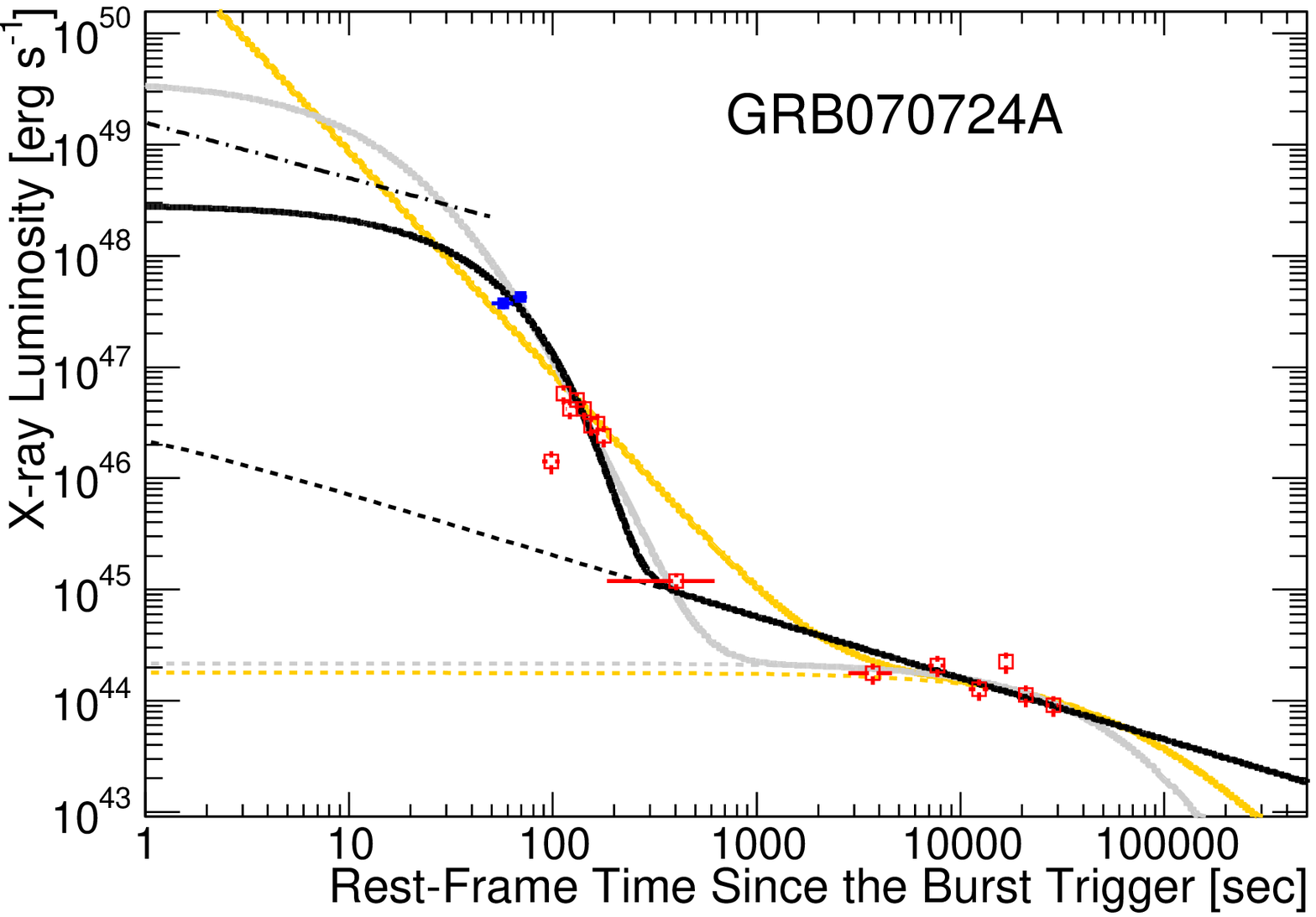}
   \includegraphics[angle=0,scale=0.28]{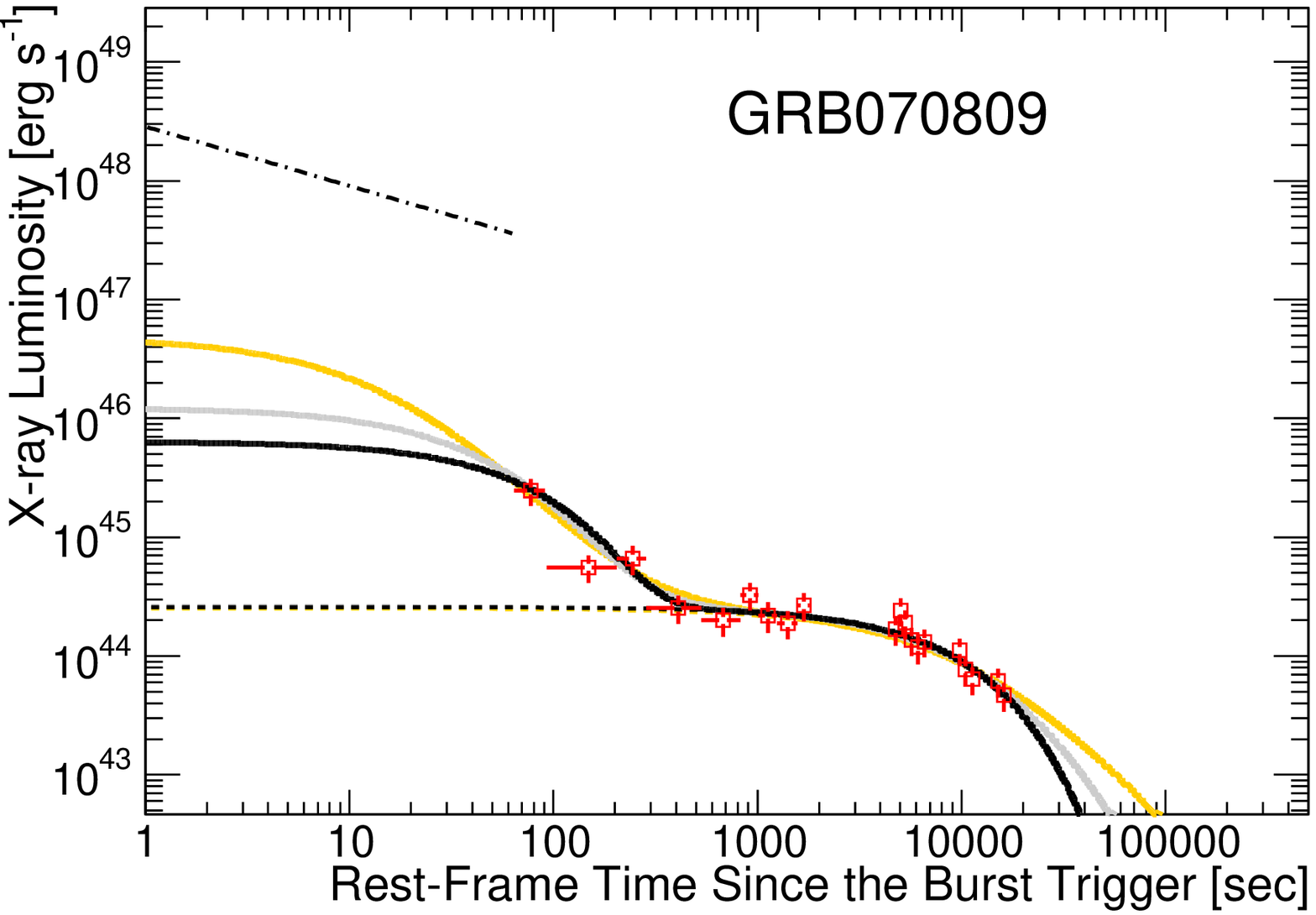}
   \includegraphics[angle=0,scale=0.28]{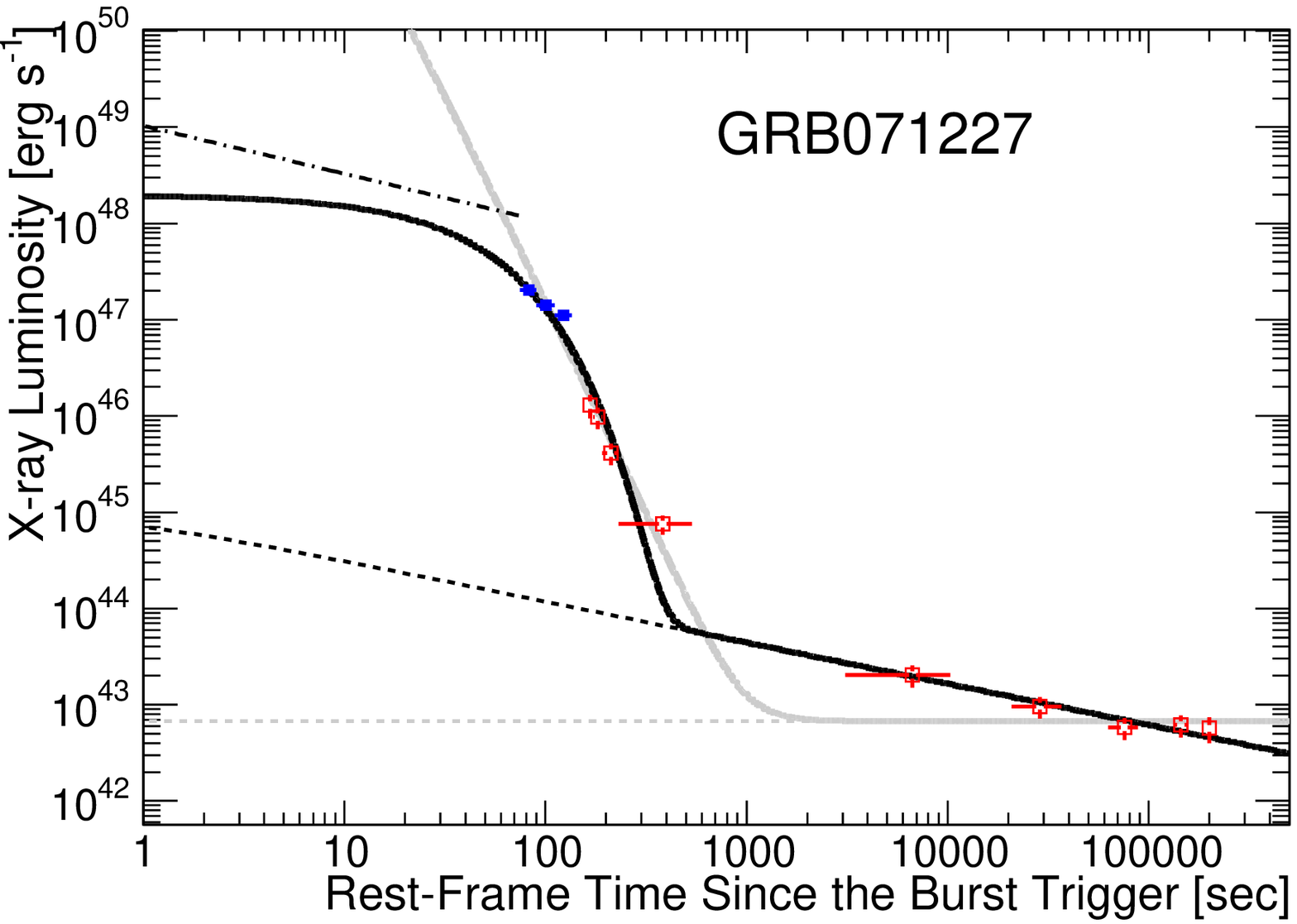}
   \includegraphics[angle=0,scale=0.28]{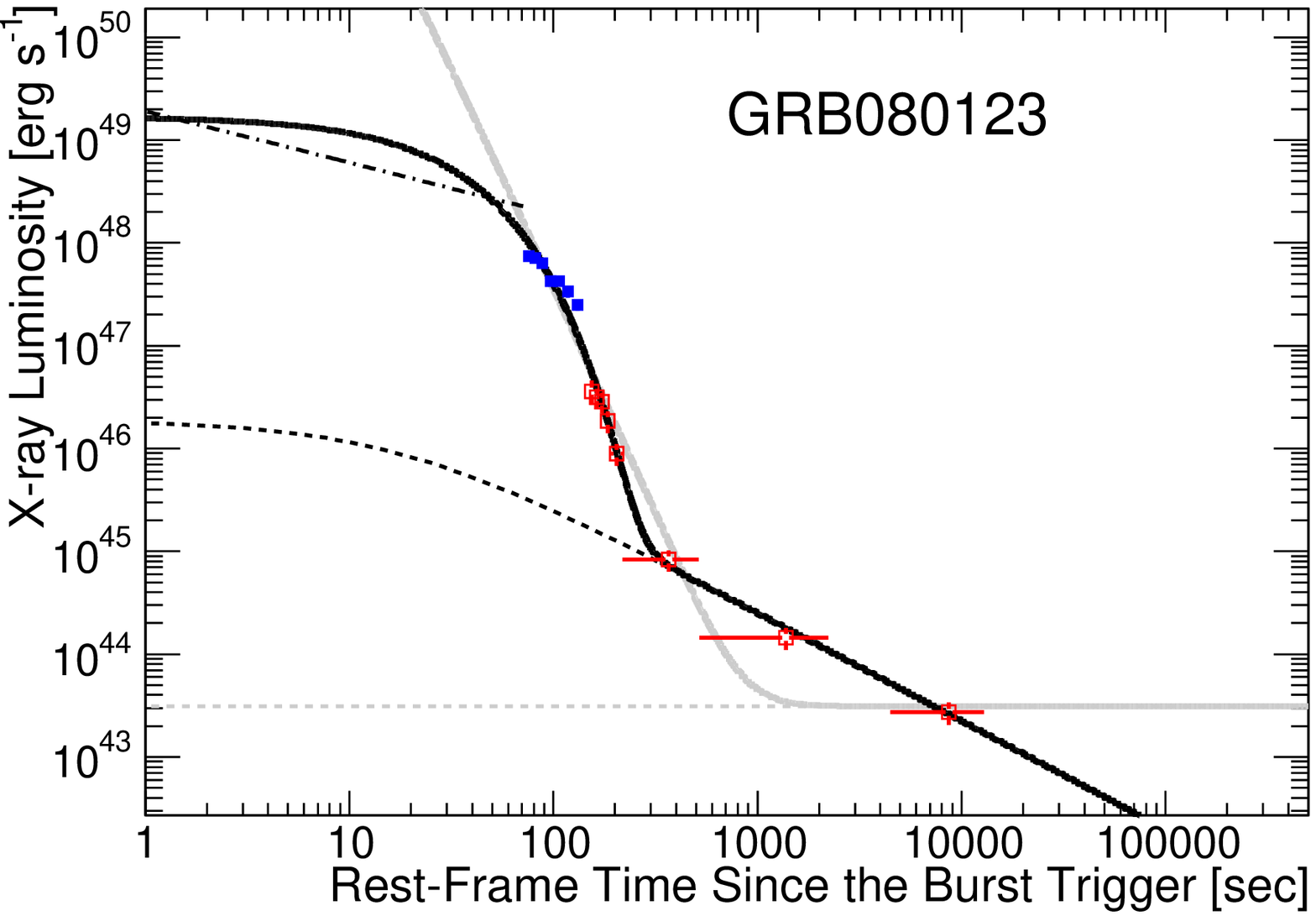}
   \includegraphics[angle=0,scale=0.28]{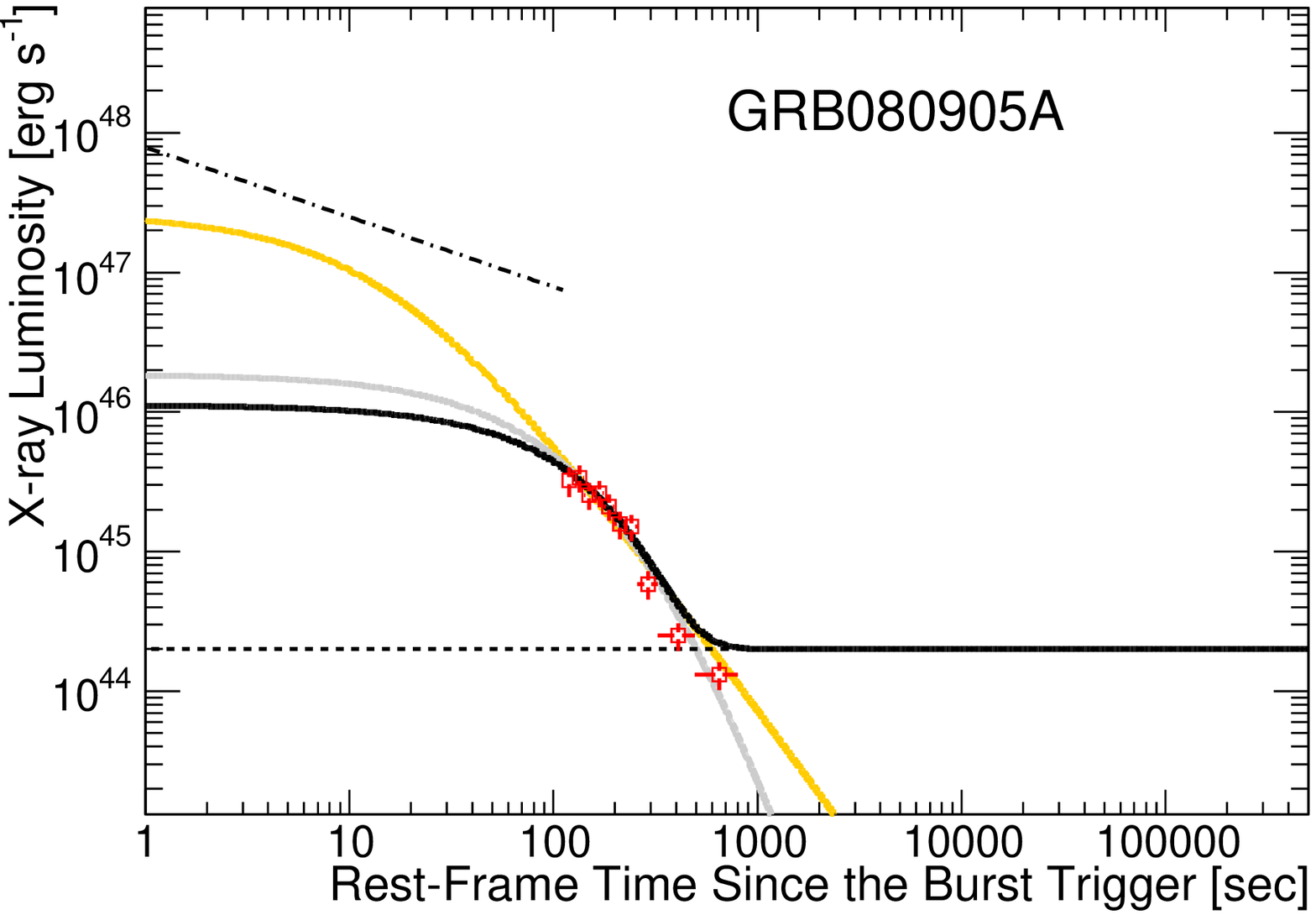}
   \includegraphics[angle=0,scale=0.28]{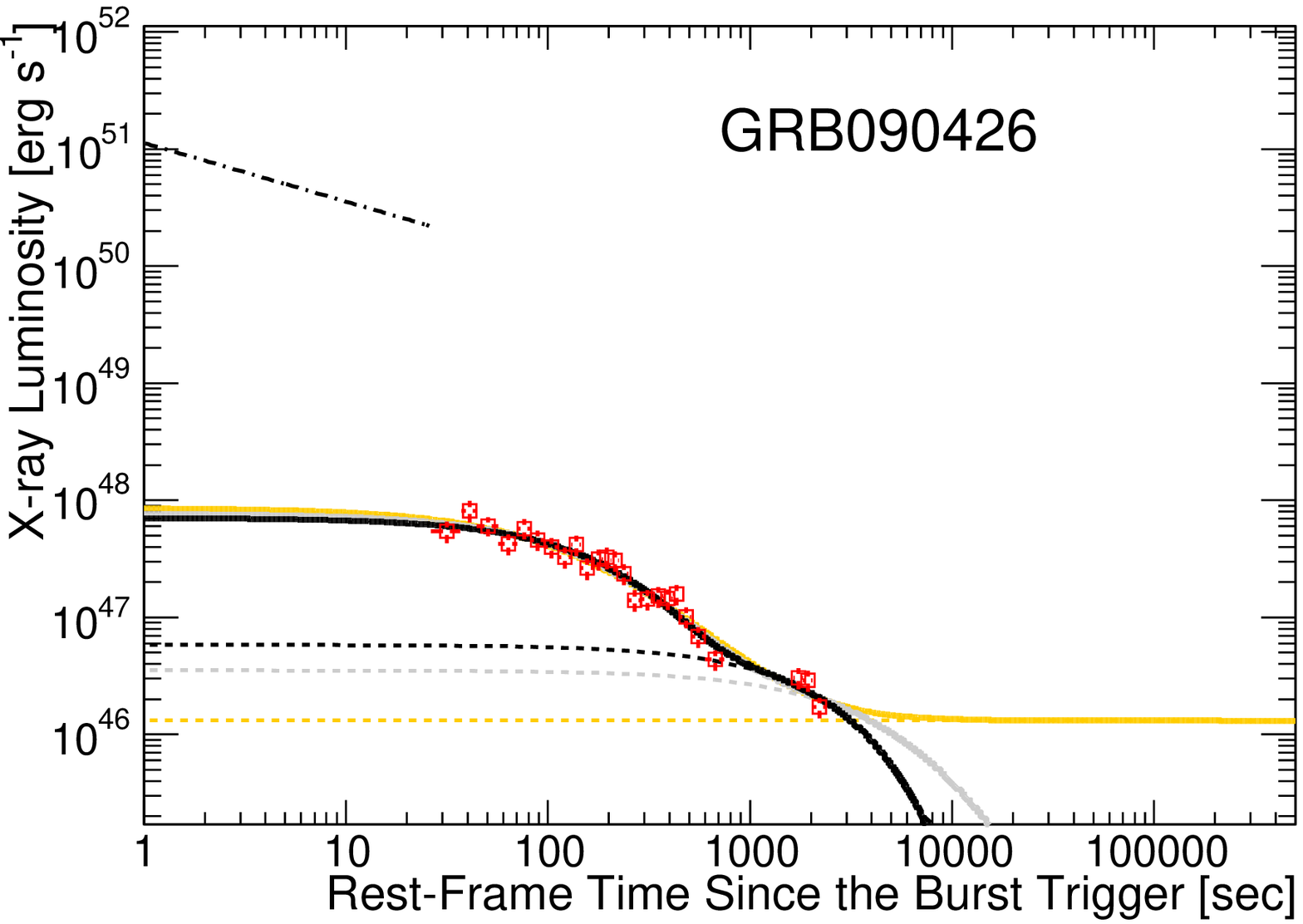}
   \includegraphics[angle=0,scale=0.28]{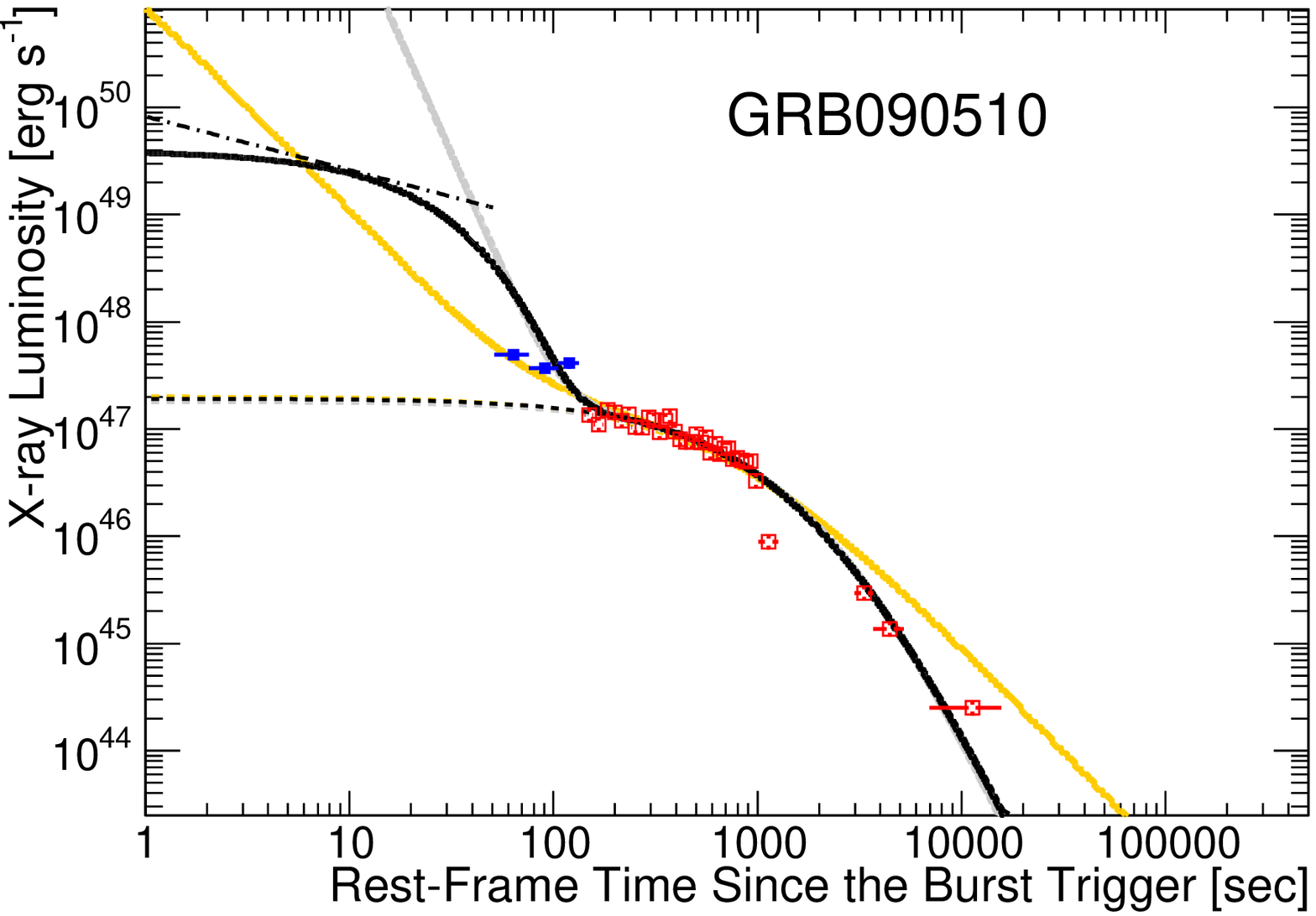}
      \includegraphics[angle=0,scale=0.28]{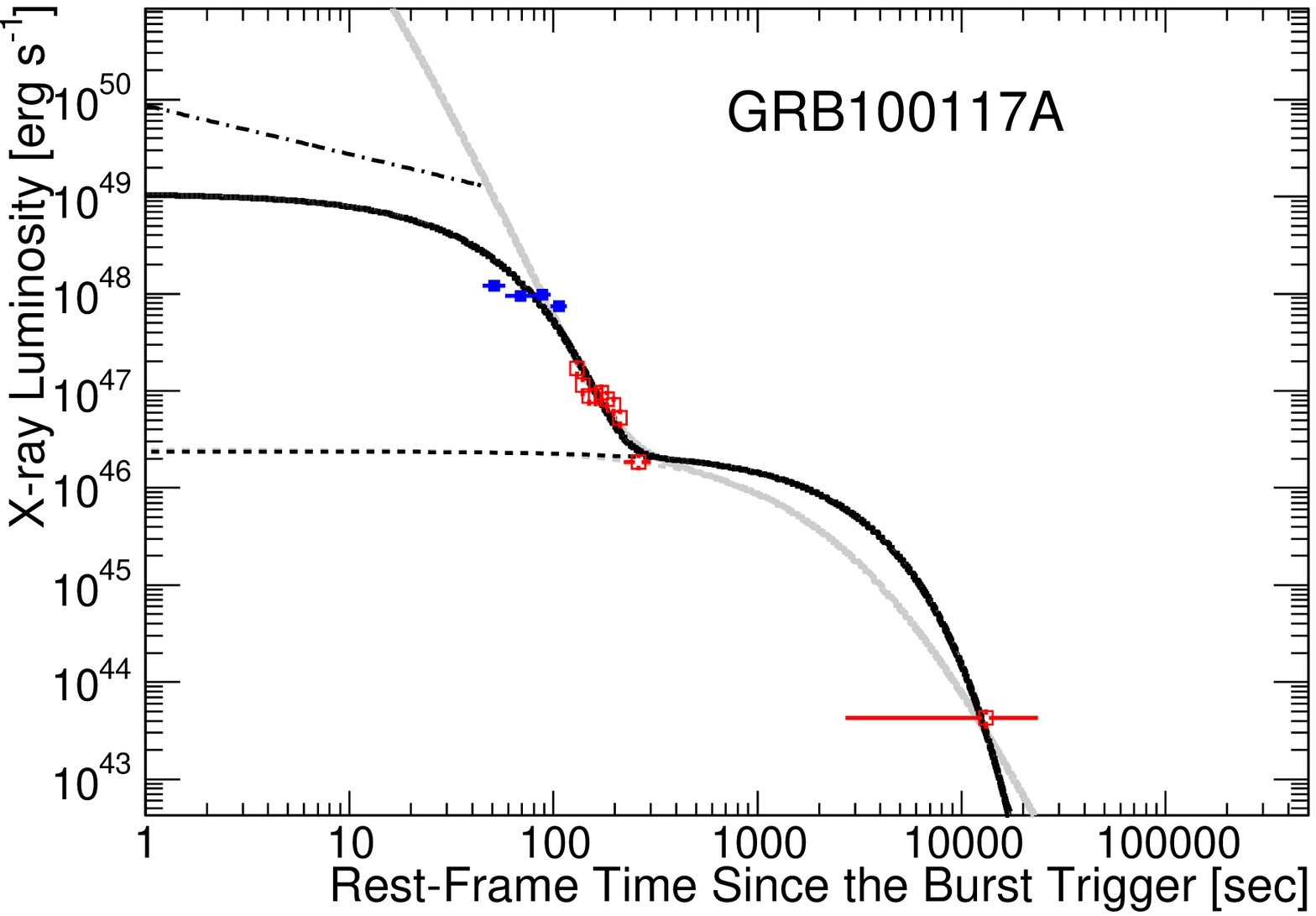}

  \end{center}
  \caption{
  The temporal history of the luminosity (in $2-10$ keV energy
  band in the rest frame) of selected SGRBs and the
  fitting curves (see Section \ref{sec:lcfit}).
  The black dash-dotted lines are the detection limits of the BAT.
  The blue-filled and red-open squares are detected with the XRT's WT
  and PC mode, respectively, and the green-open circles are obtained
  with the BAT observation.
  The black and gray solid lines show the best-fitted EXP and PL(BH) models,
  respectively and the orange solid line drawn for only the events with
  $\chi^2_{\nu}$ of $<7$ (see Table \ref{table:results}) shows the best-fitted
  PL(MG) model. The black, gray, and orange dashed lines show the plateau
  emission components.
  } 
   \label{fig:lcfit}
 \end{figure}
 
 \addtocounter{figure}{-1}
 \begin{figure}[tbp]
   \begin{center}
   \includegraphics[angle=0,scale=0.28]{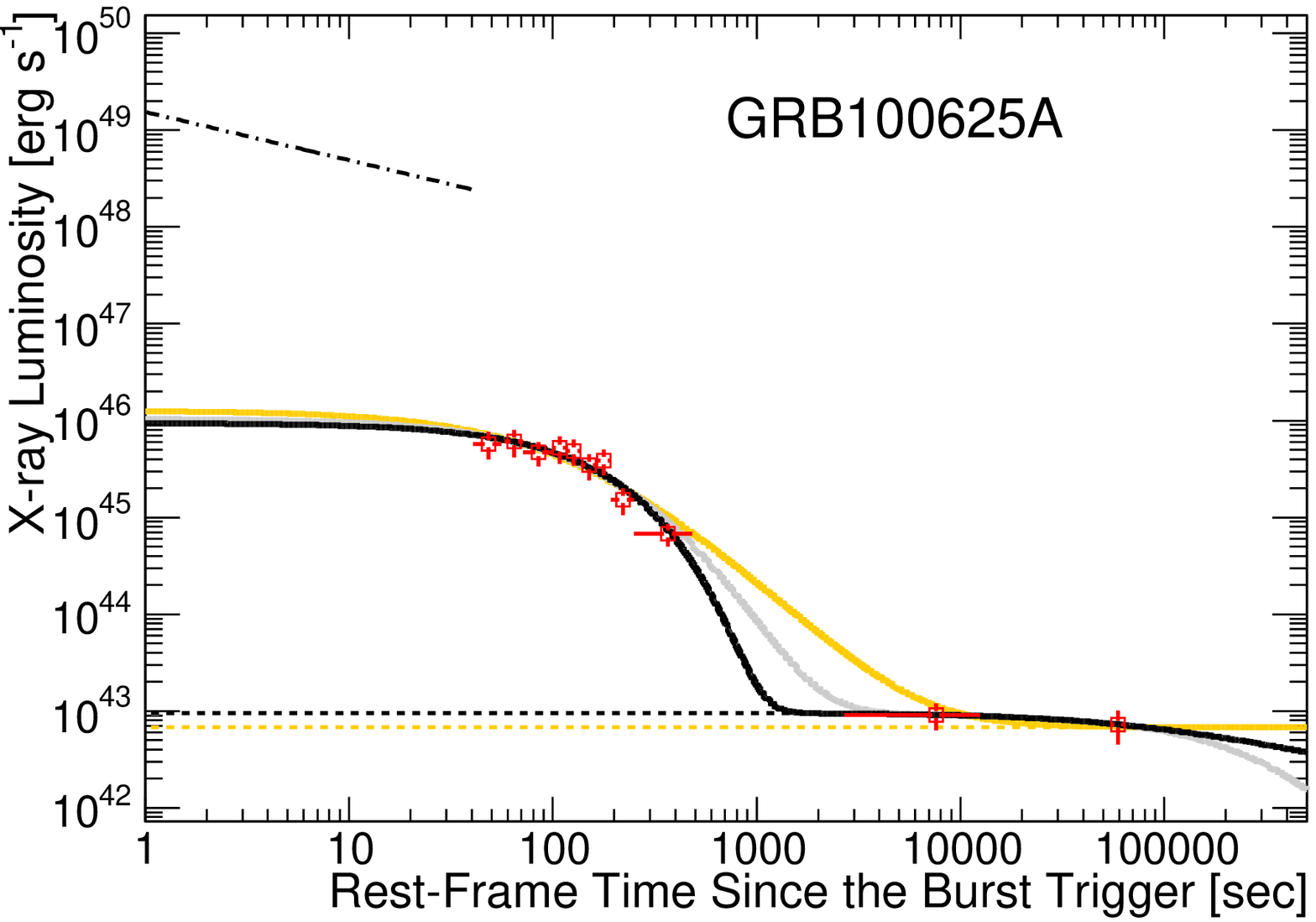}
   \includegraphics[angle=0,scale=0.28]{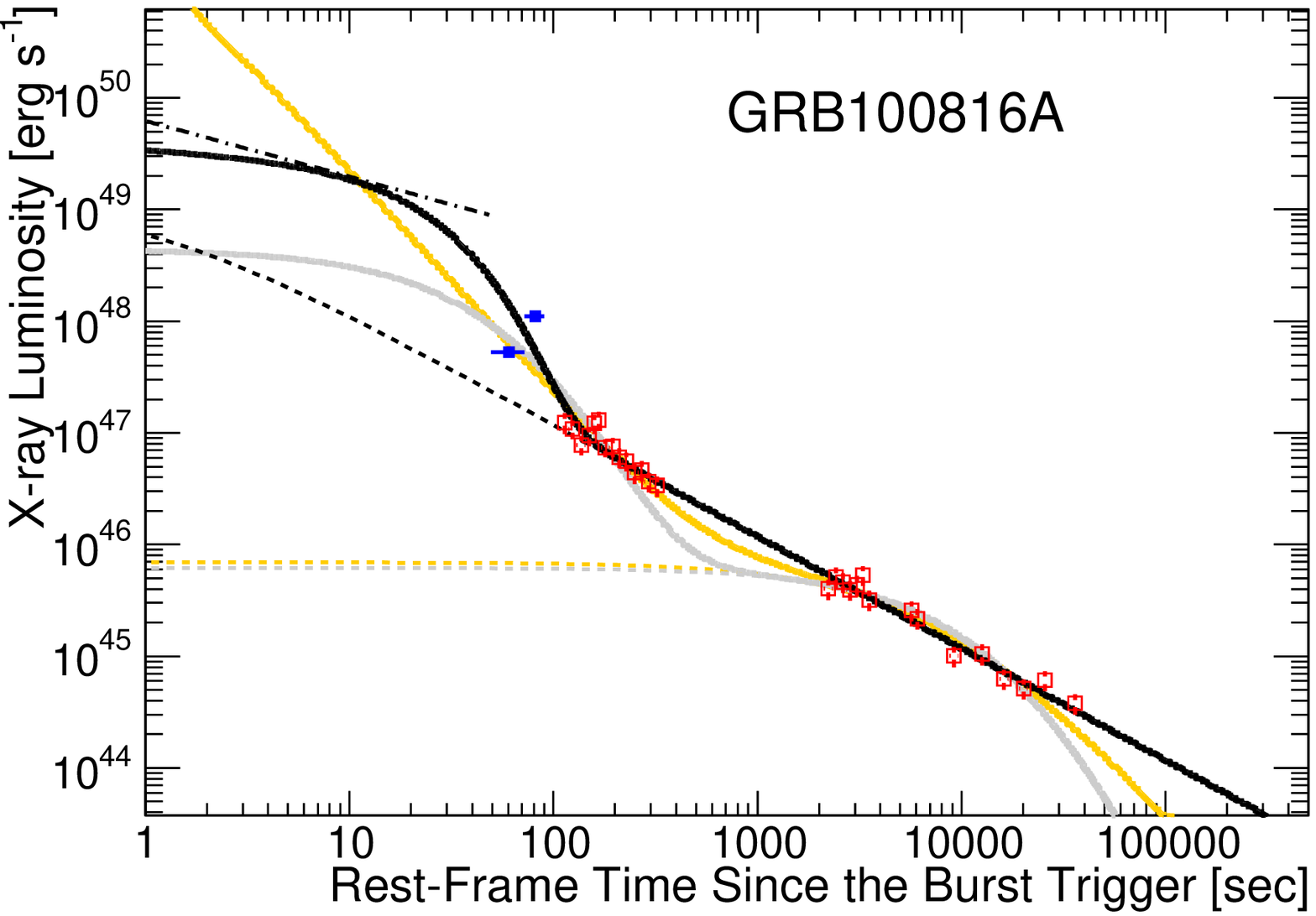}
   \includegraphics[angle=0,scale=0.28]{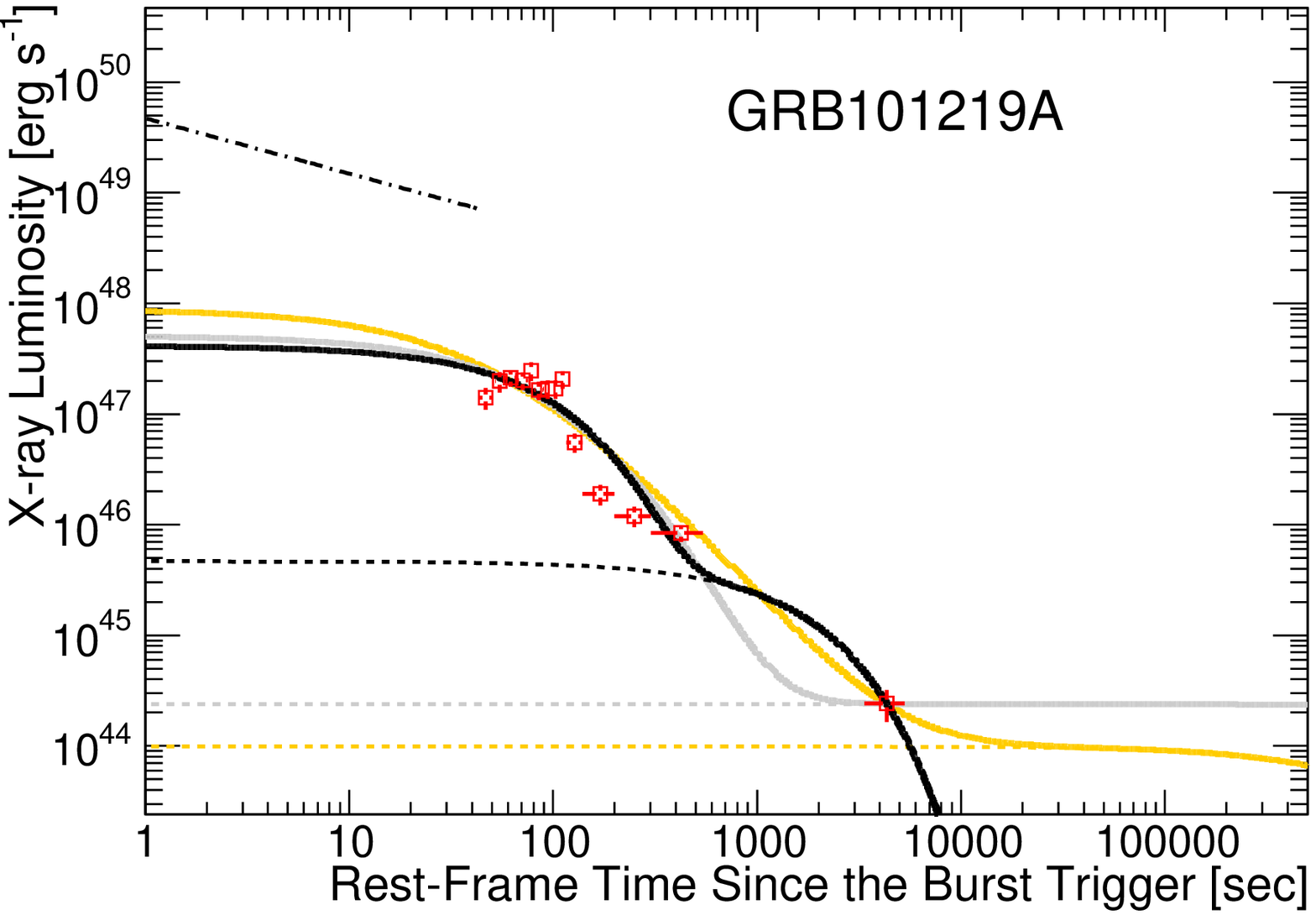}
   \includegraphics[angle=0,scale=0.28]{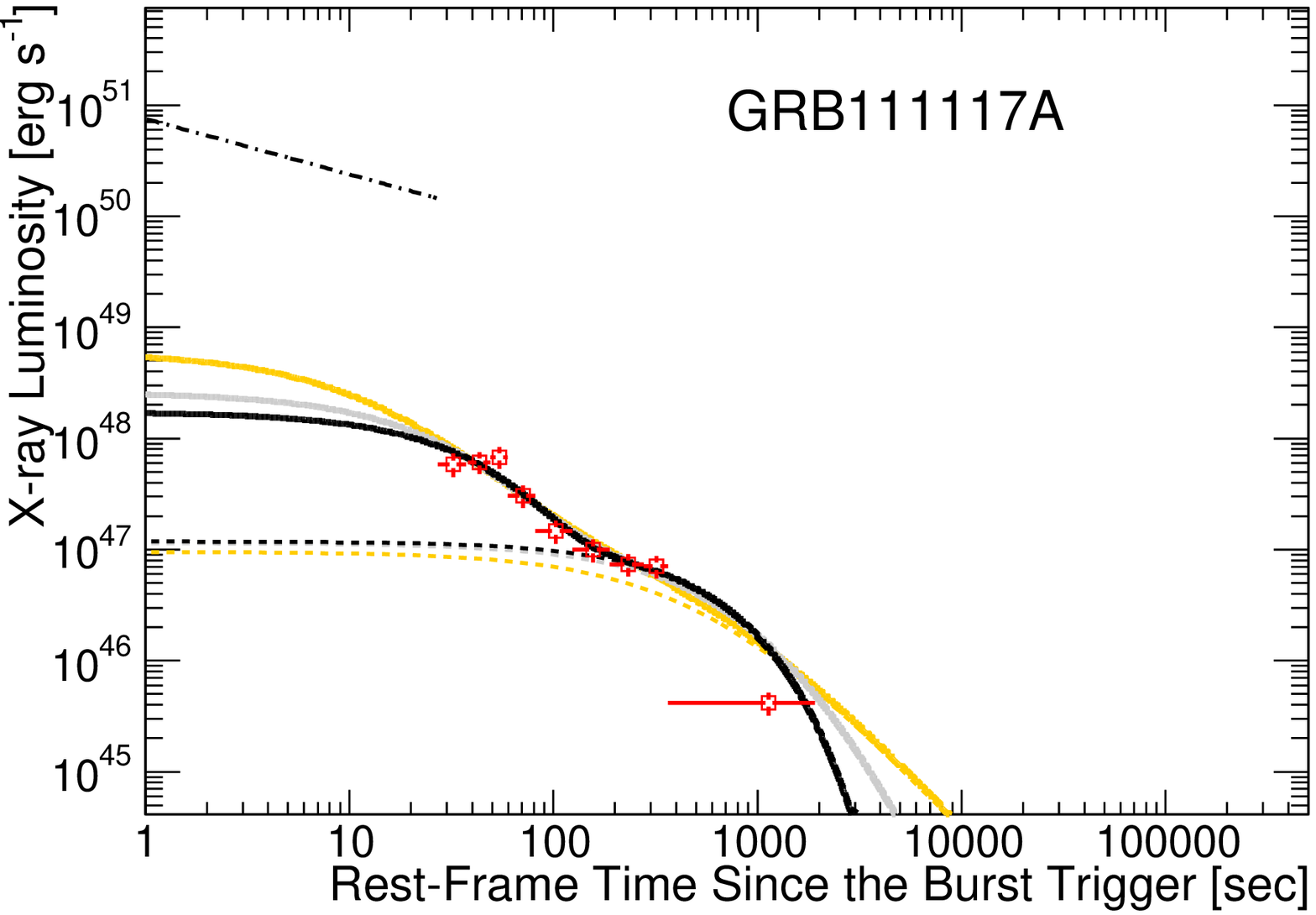}
   \includegraphics[angle=0,scale=0.28]{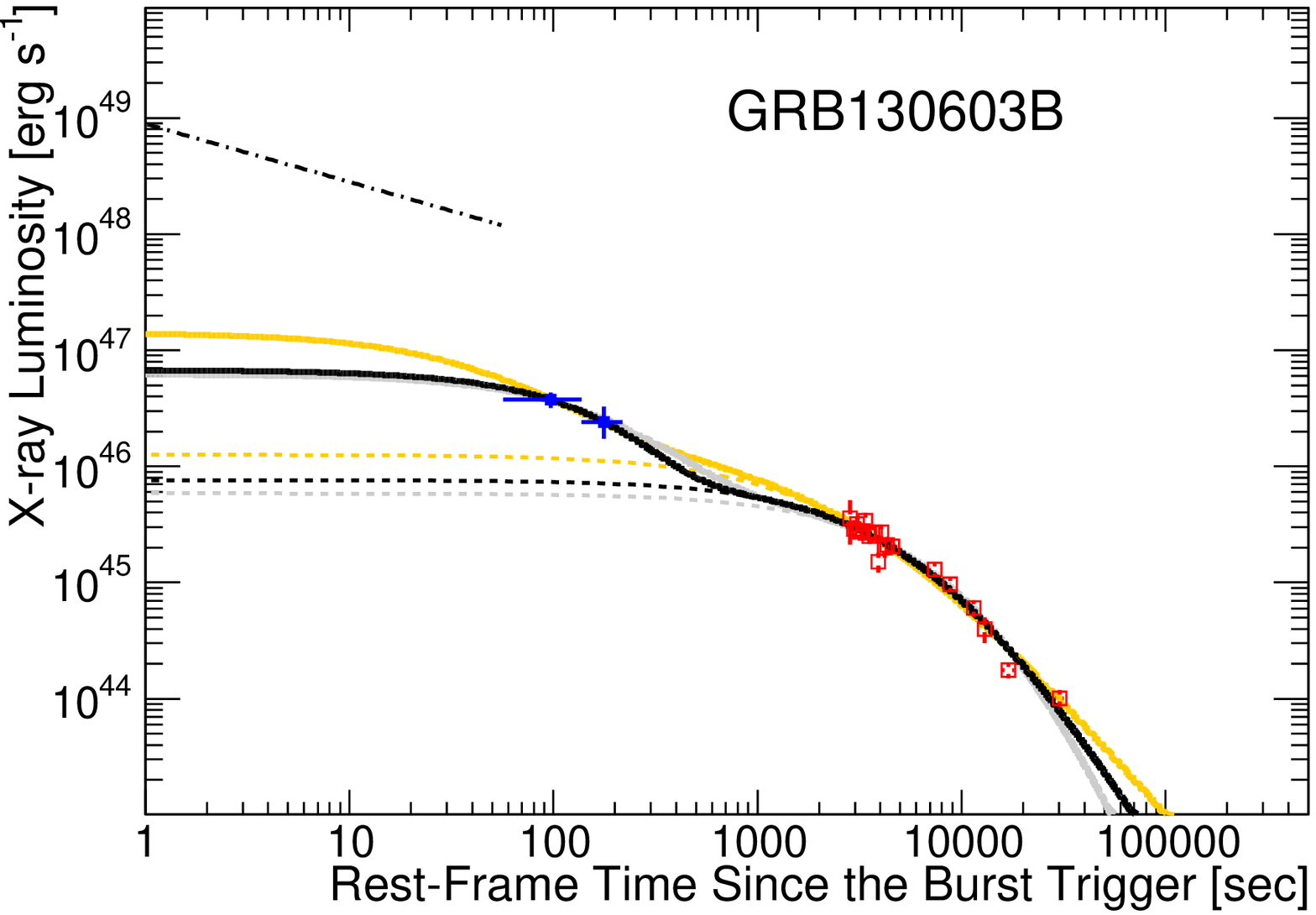}
   \includegraphics[angle=0,scale=0.28]{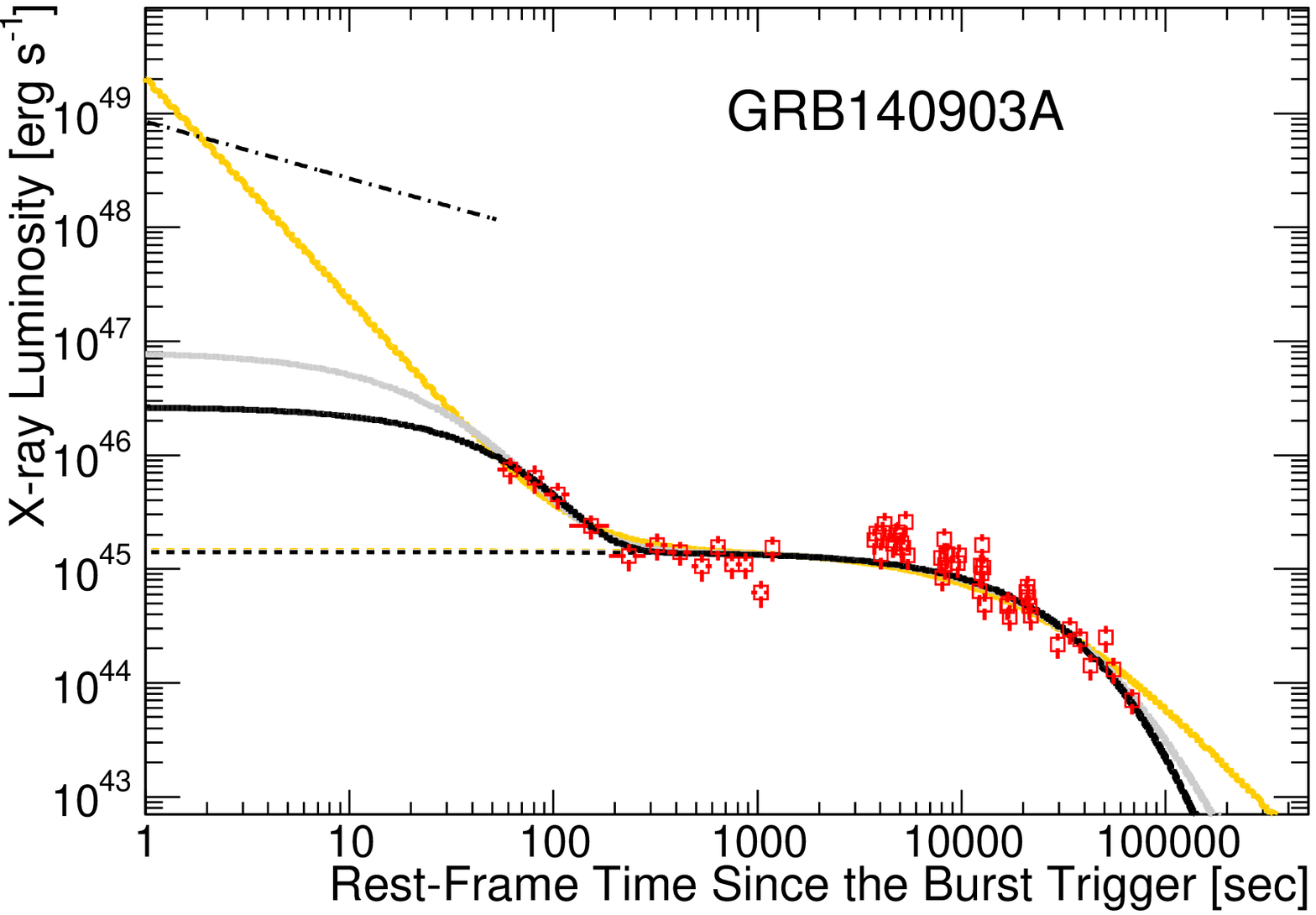}
   \includegraphics[angle=0,scale=0.28]{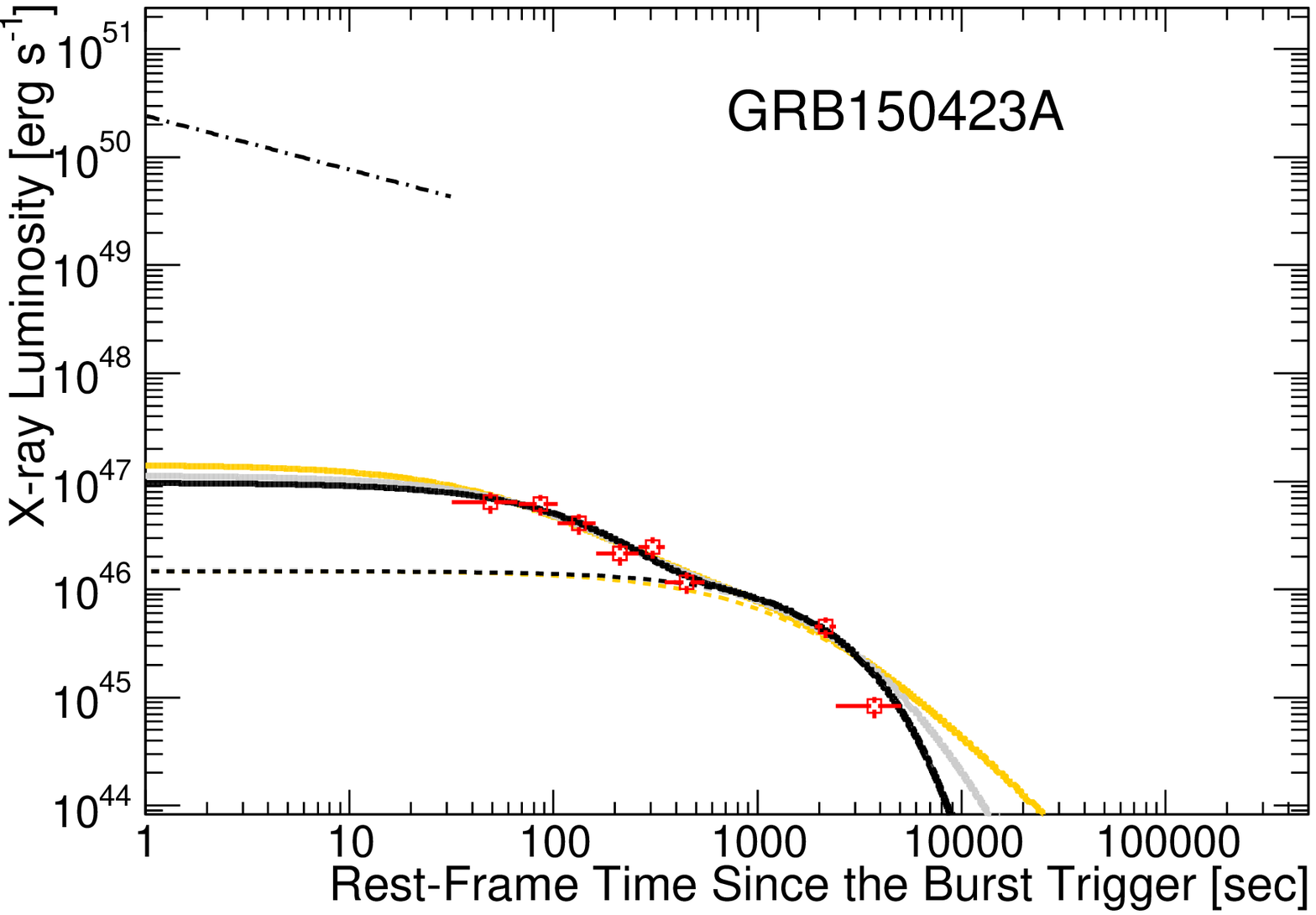}
   \includegraphics[angle=0,scale=0.28]{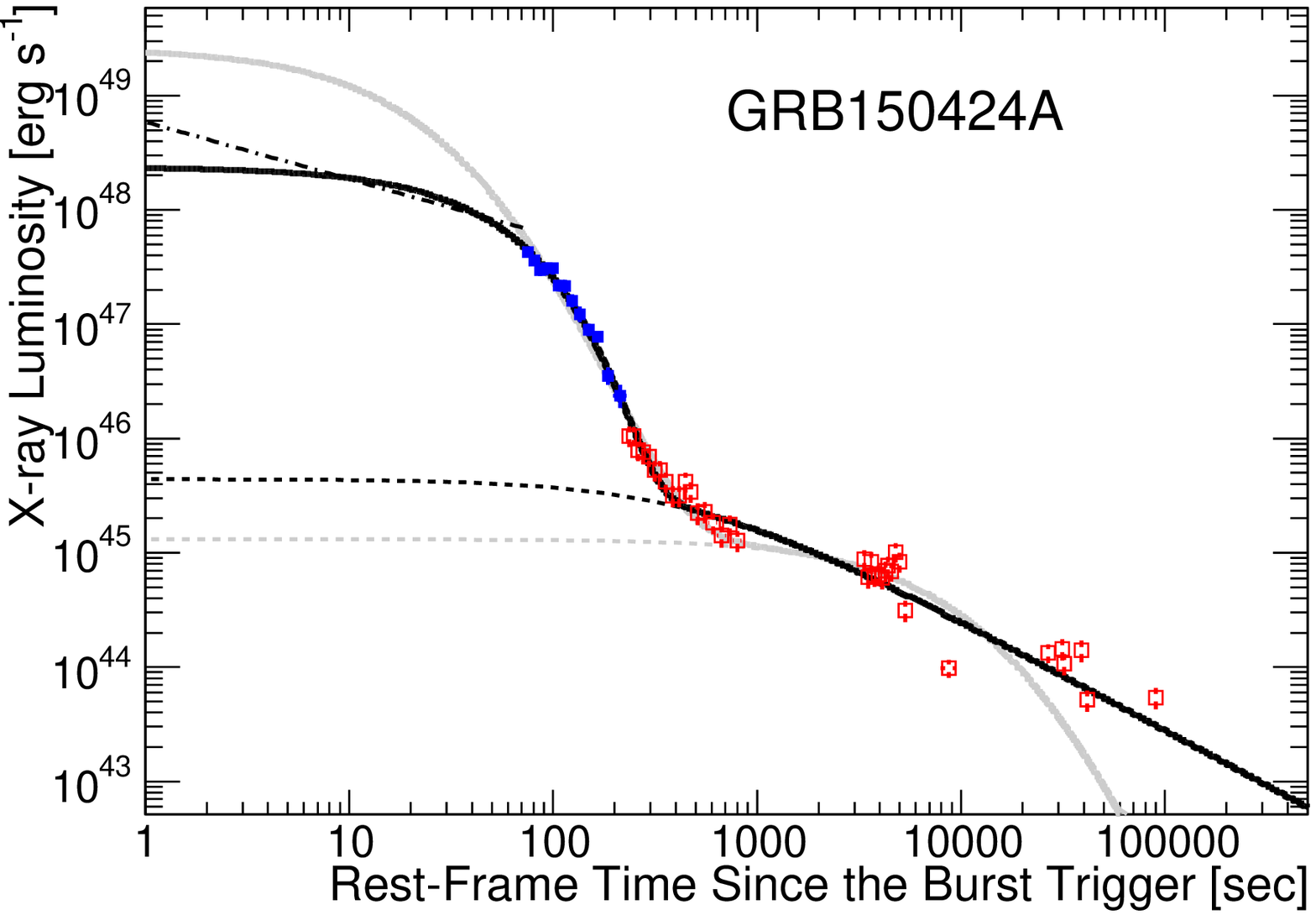}
   \includegraphics[angle=0,scale=0.28]{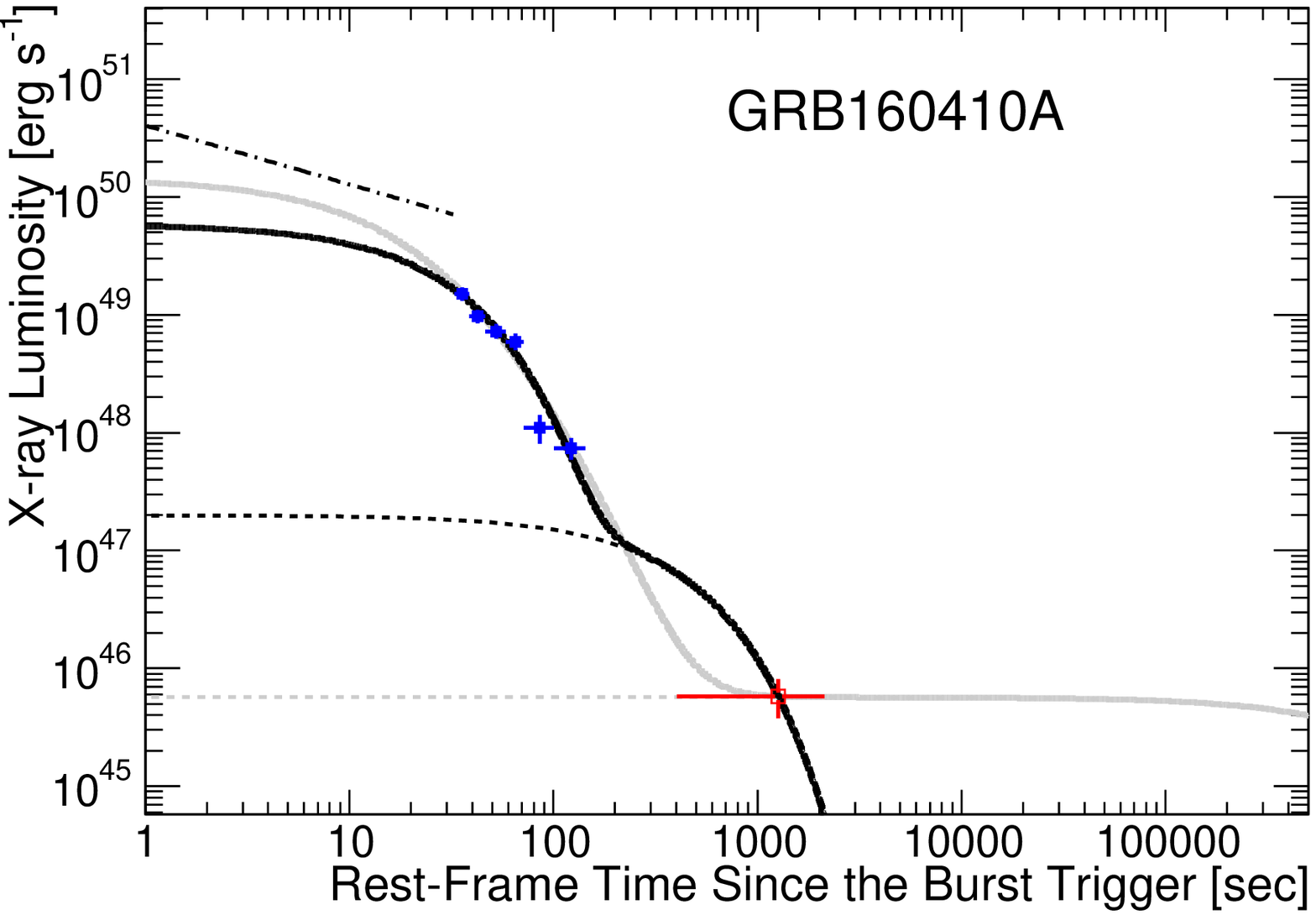}
   \includegraphics[angle=0,scale=0.28]{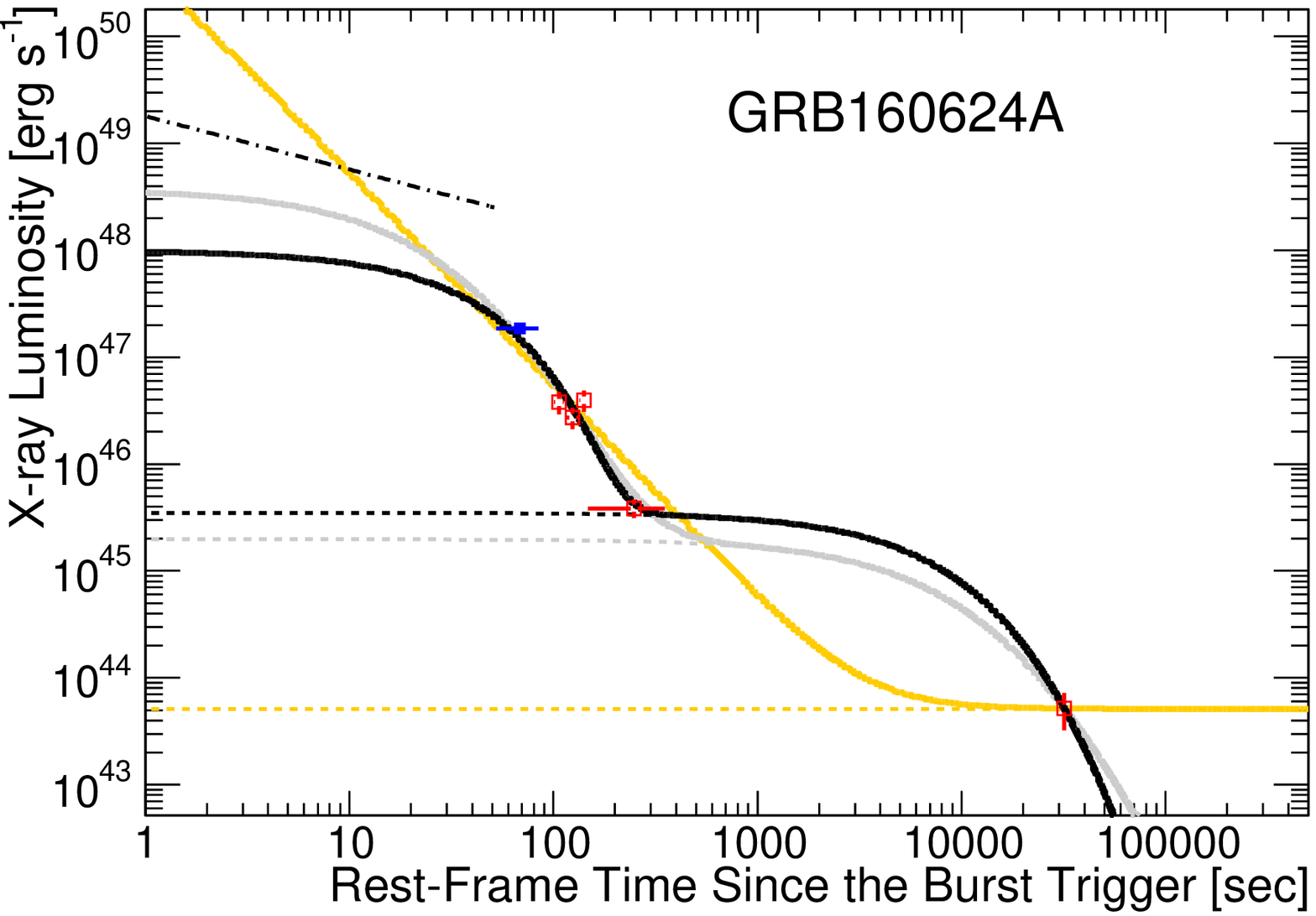}
   \includegraphics[angle=0,scale=0.28]{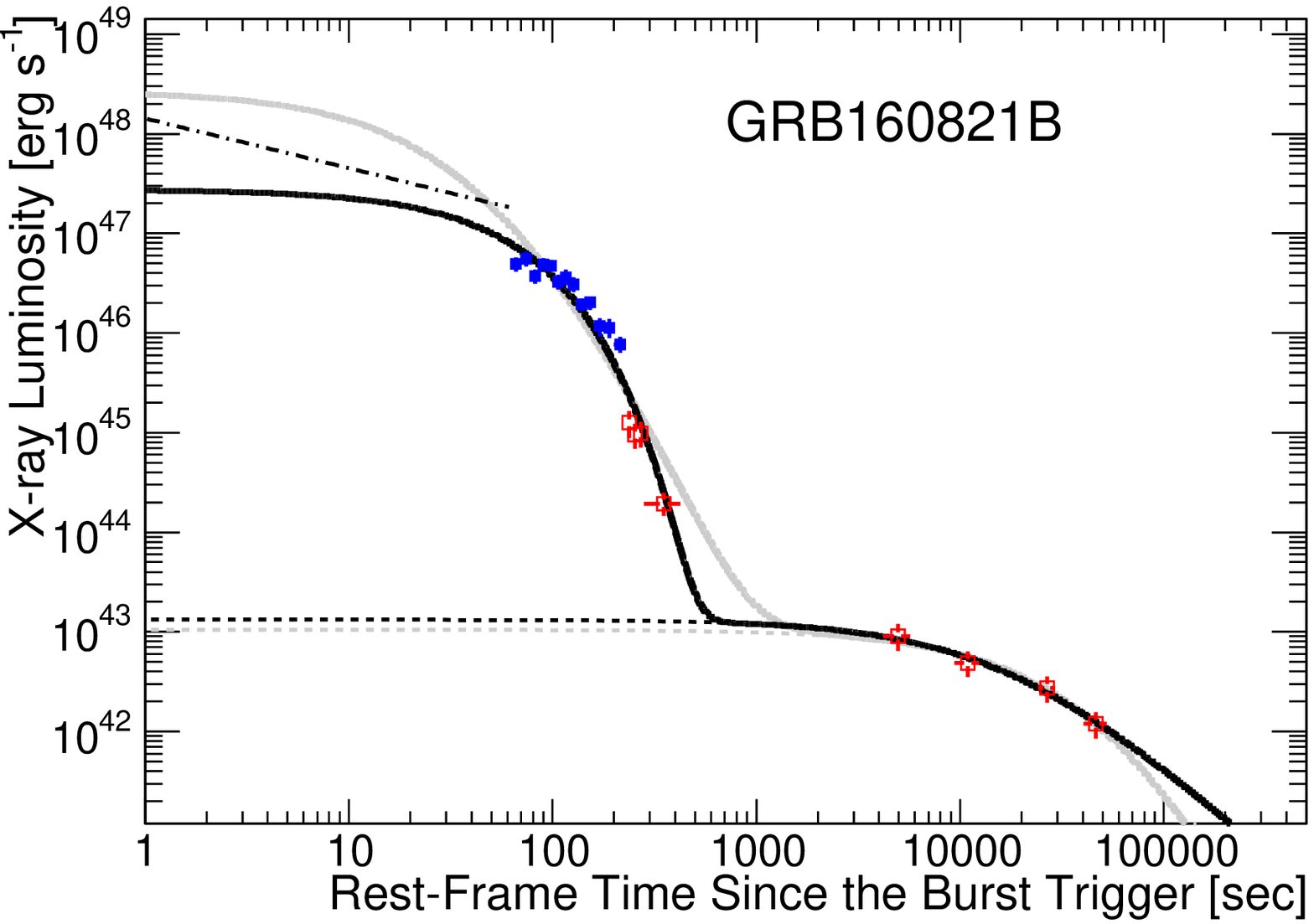}
   \caption{
   Continued.
   } 
   \label{fig:lcfit}
   \end{center}
 \end{figure}
 
\begin{figure}
 \begin{center}
  \includegraphics[angle=0,scale=0.8]{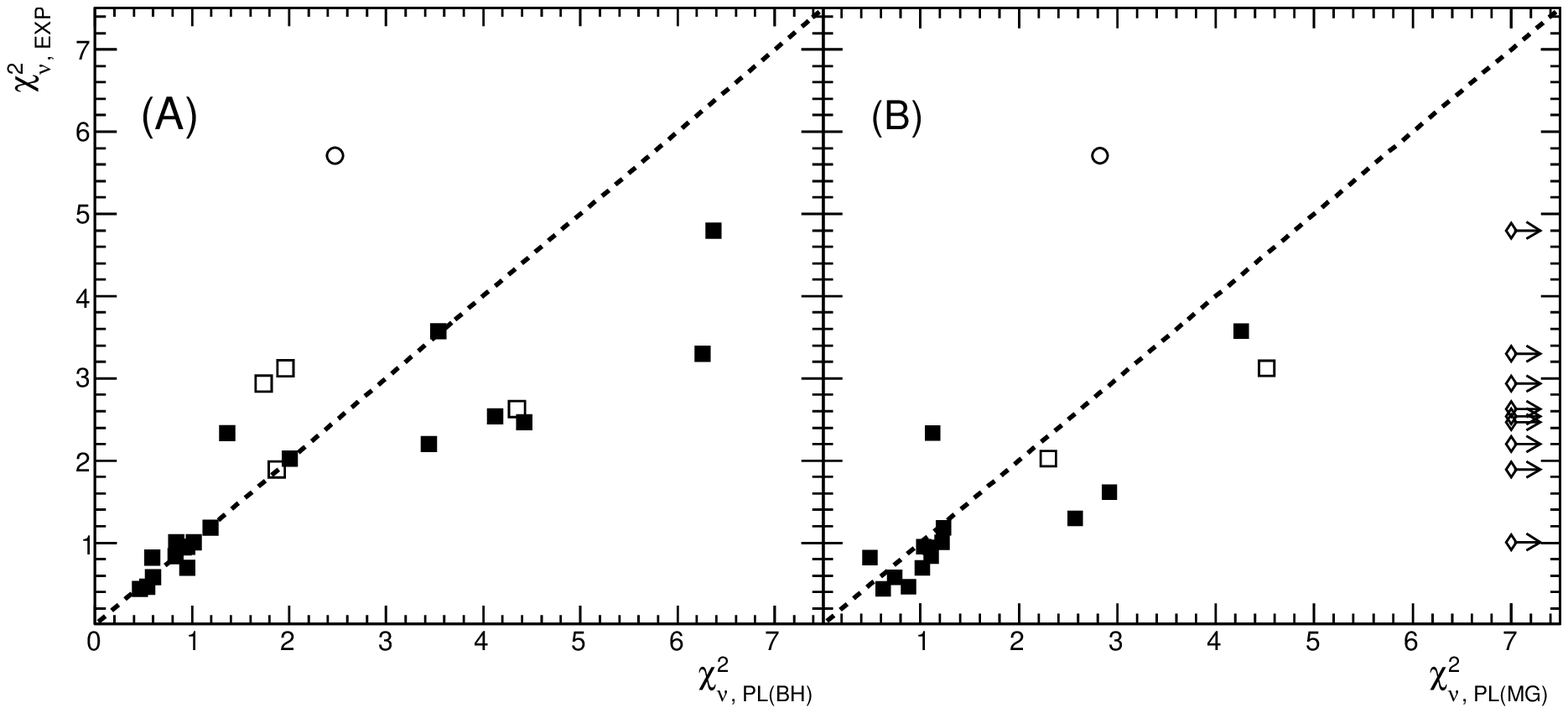}
  \caption{
  Scatter plot of the reduced $\chi^2$ of the EXP and PL(BH) models in
  the (A) panel and of the EXP and PL(MG) models in the (B) panel. 
  The dotted lines in the (A) and (B) panel correspond to $\chi^2_{\nu,
  {\rm EXP}} = \chi^2_{\nu, {\rm PL(BH)}}$ and $\chi^2_{\nu,
  {\rm EXP}} = \chi^2_{\nu, {\rm PL(MG)}}$, respectively.
  The opened squares correspond to the events whose PL(BH)
  model curve exceeds the BAT detection limit as shown in Figure
  \ref{fig:lcfit}. The opened circle denotes the data of GRB~160624A
  whose degrees of freedom of the EXP and PL(BH) models obtained by the
  light curve fitting are 1 and 2, respectively (see also Table
  \ref{table:results}).
  The opened diamonds in the (B) panel denote the data of the events
  with $\chi^{2}_{\nu, {\rm PL(MG)}}>7$.
  In these figures, the data of GRB~090510 and 100816A are not plotted.
  }  
  \label{fig:chi}
 \end{center}
\end{figure}

\begin{figure}
 \begin{center}
  \includegraphics[angle=0,scale=0.6]{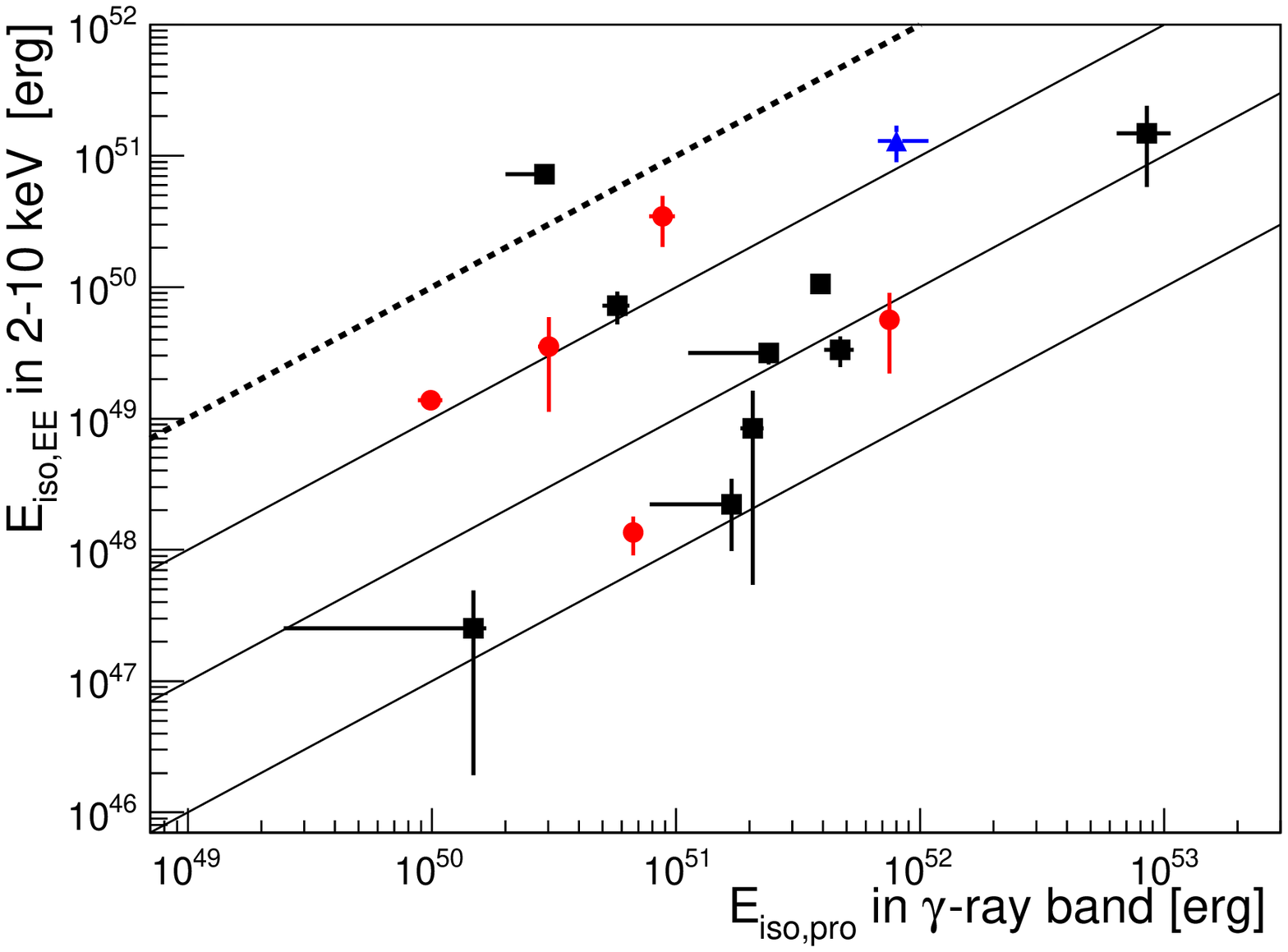}
  \caption{
  Scatter plot of the isotropic energies of the prompt emission $E_{{\rm
  iso, pro}}$ in the $\gamma$-ray band and extended emission $E_{{\rm
  iso, EE}}$ in the 2 -- 10 keV energy band.
  The black squares, red circles and blue triangle mean the events whose
  prompt emissions are detected with the WIND/Konus, {\it Fermi}/GBM, and
  {\it Suzaku}/WAM, respectively. 
  As listed in \ref{table:prompt}, $E_{{\rm iso, pro}}$ is measured in
  different energy band, but the peak energy $E_{\rm peak}$ is well
  covered by each detector. Therefore, $E_{{\rm iso, pro}}$ can be
  recognized as the nearly bolometric energy.
  $E_{{\rm iso,EE}}$ equals to $L_{EE}\times\tau_{{\rm EE}}$ as Equation
  \ref{eq:exp}. The dashed line corresponds to $E_{{\rm iso,
  EE}}/E_{{\rm iso, pro}} = 1$ and the three solid lines correspond to
  $E_{{\rm iso, EE}}/E_{{\rm iso, pro}} = 0.1,\  0.01, {\rm and}\ 0.001$
  from top to bottom, respectively. 
  }  

  \label{fig:eiso}
 \end{center}
\end{figure}
\begin{figure}
 \begin{center}
  \includegraphics[angle=0,scale=0.5]{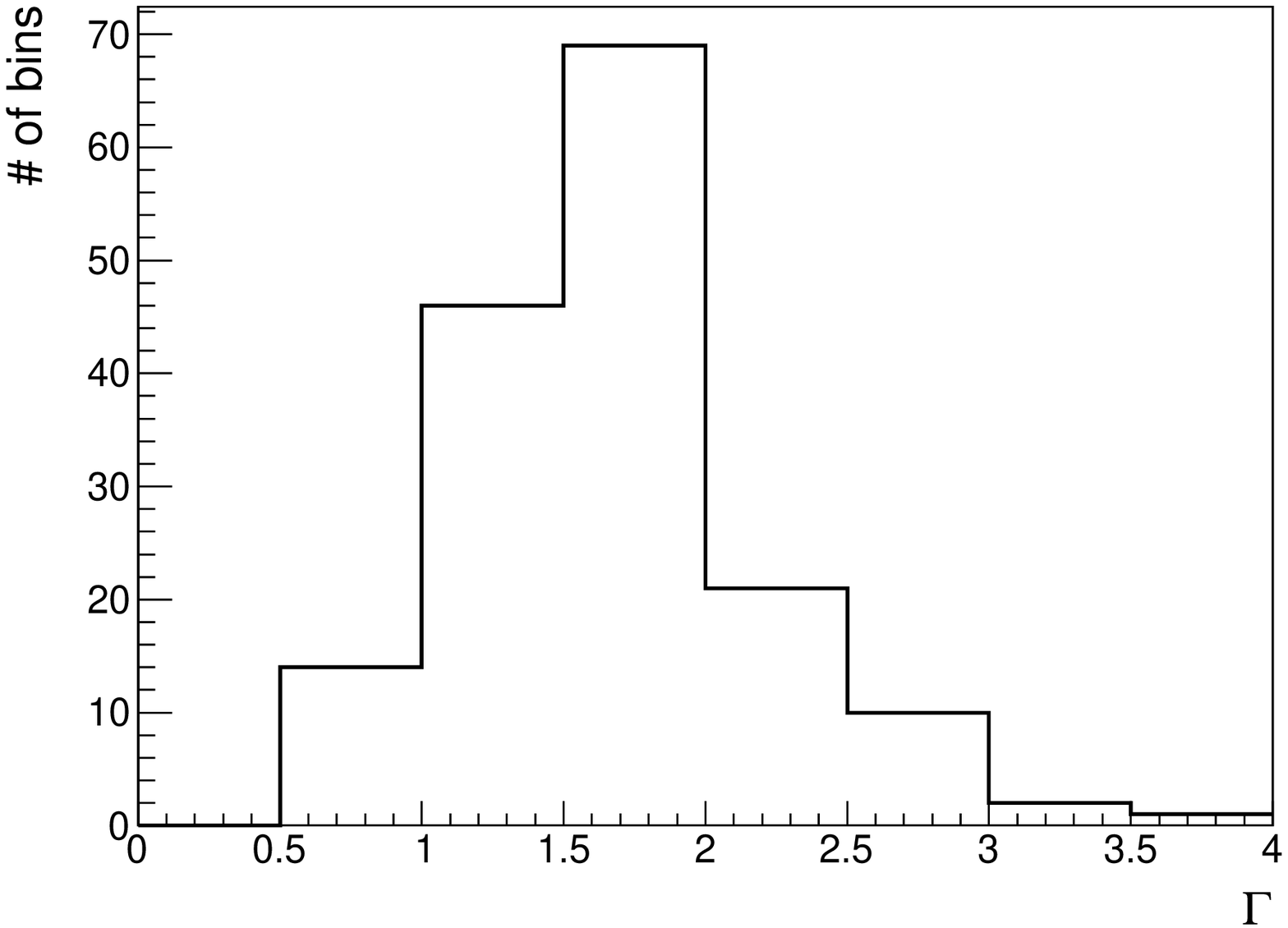}
  \caption{
  A histogram of the photon index parameters $\Gamma$ obtained by
  performing time-resolved analysis for the data observed with WT mode
  of the XRT.
  The mean value and the standard deviation of the histogram are $1.7$
  and $0.5$, respectively. 
  }  
  \label{fig:gamma}
 \end{center}
\end{figure}
 
 \begin{figure}
  \begin{center}
   \includegraphics[angle=0,scale=0.5]{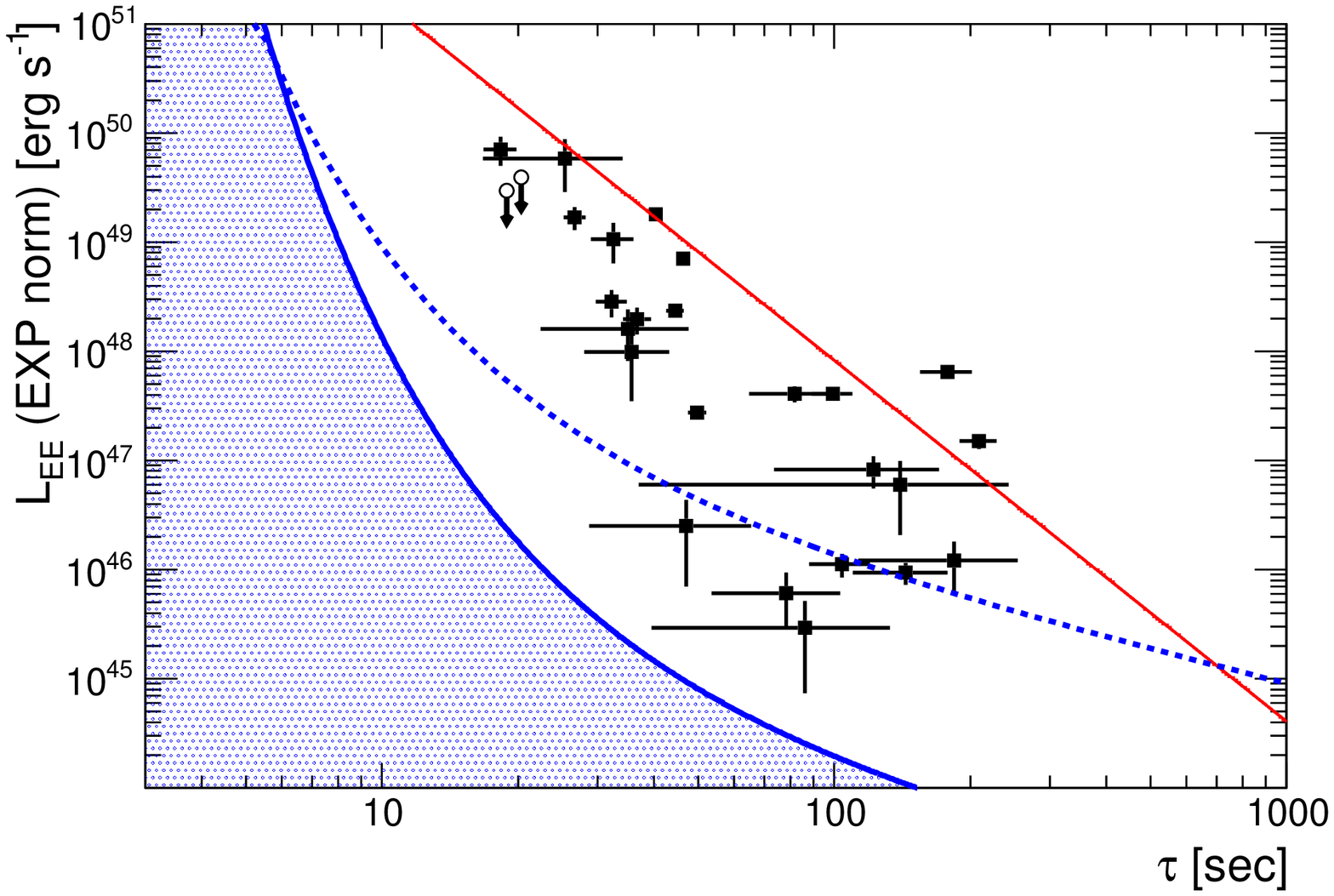}
   \caption{
   Scatter plot of $L_{{\rm EE}} - \tau_{{\rm EE}}$. The open circles with
   the down arrow correspond to the data of GRB~090510 and GRB~100816A
   suppressed by the BAT detection limit, respectively.
   The blue solid line shows the sensitivity limit of $L_{{\rm limit,
   XRT}}(\tau_{{\rm EE}})$ considering that SGRBs occur at $z=0.10$ and
   the events on the blue shaded area are undetectable for the {\it
   Swift}/XRT (see text for details). 
   The blue dashed line exhibits the same formula of $z=0.72$, which is an
   averaged redshift value of observed SGRBs \citep{kisaka2017}.
   The solid red line denotes the best fit of power-law function adopted
   to the plot data (see Section\ref{subsec:l-tau}).
   } 
   \label{fig:ltau}
  \end{center} 
 \end{figure}

\end{document}